\documentclass[12pt]{article}
\usepackage{graphicx}
\usepackage{epsfig}
\usepackage{epstopdf}
\DeclareGraphicsExtensions{.pdf,.eps,.png,.jpg,.mps}

\def\lsim{\mathrel{\rlap{\lower3pt\hbox{\hskip0pt$\sim$}}
     \raise1pt\hbox{$<$}}}         
\def\gsim{\mathrel{\rlap{\lower4pt\hbox{\hskip1pt$\sim$}}
     \raise1pt\hbox{$>$}}}         

\usepackage{amsmath}
\usepackage{amsfonts}

\begin{document}
\begin{titlepage}

\centerline{\Large \bf Notes on Alpha Stream Optimization}
\medskip

\centerline{Zura Kakushadze$^\S$$^\dag$\footnote{\, Zura Kakushadze, Ph.D., is the President of Quantigic$^\circledR$ Solutions LLC,
and a Full Professor at Free University of Tbilisi. Email: \tt zura@quantigic.com}}
\bigskip

\centerline{\em $^\S$ Quantigic$^\circledR$ Solutions LLC}
\centerline{\em 1127 High Ridge Road \#135, Stamford, CT 06905\,\,\footnote{\, DISCLAIMER: This address is used by the corresponding author for no
purpose other than to indicate his professional affiliation as is customary in
publications. In particular, the contents of this paper
are not intended as an investment, legal, tax or any other such advice,
and in no way represent views of Quantigic® Solutions LLC,
the website \underline{www.quantigic.com} or any of their other affiliates.
}}
\centerline{\em $^\dag$ Free University of Tbilisi, Business School \& School of Physics}
\centerline{\em 240, David Agmashenebeli Alley, Tbilisi, 0159, Georgia}
\medskip
\centerline{(June 4, 2014; revised: March 18, 2015)}

\bigskip
\medskip

\begin{abstract}
{}In these notes we discuss investment allocation to multiple alpha streams traded on the same execution platform, including when trades are crossed internally resulting in turnover reduction. We discuss approaches to alpha weight optimization where one maximizes P\&L subject to bounds on volatility (or Sharpe ratio). The presence of negative alpha weights, which are allowed when alpha streams are traded on the same execution platform, complicates the optimization problem. By using factor model approach to alpha covariance matrix, the original optimization problem can be viewed as a 1-dimensional root searching problem plus an optimization problem that requires a finite number of iterations. We discuss this approach without costs and with linear costs, and also with nonlinear costs in a certain approximation, which makes the allocation problem tractable without forgoing nonlinear portfolio capacity bound effects.
\end{abstract}
\end{titlepage}

\newpage

\section{Introduction and Summary}

{}Nowadays, technology allows to combine a large number of hedge fund alpha streams on the same trading platform.\footnote{\, For a partial list of hedge fund literature, see, {\em e.g.}, \cite{HF1}-\cite{HF20} and references therein.} Not only does this yield diversification, but also reduces transaction costs by crossing trades between different alpha streams.\footnote{\, For a recent discussion, see \cite{OD}.} Given a set of alpha streams, one has to decide how to allocate funds into these alphas, {\em i.e.}, the weights with which to combine them.\footnote{\, For a partial list of portfolio optimization and related literature, see, {\em e.g.}, \cite{PO1}-\cite{PO35} and references therein.} This is an old dilemma between greed and prudence -- does one maximize the P\&L, does one minimize the volatility, does one do something in between, or does one simply split the difference and maximize the Sharpe ratio instead? It all depends on one's risk tolerance. For instance, one could say, I want to maximize my P\&L, but I also want my volatility to be capped, or the Sharpe ratio to be bounded from below, {\em etc}.

{}The purpose of these notes is to discuss aspects of optimizing weights for alpha streams. When we have a large number $N$ of alphas $\alpha_i$, this optimization is an $N$-dimensional problem and does not always have a simple solution. One way to tackle it is to try to essentially reduce it to a 1-dimensional problem, by considering simpler problems of maximizing P\&L for fixed volatility or Sharpe ratio, or by maximizing (minimizing) Sharpe ratio (volatility) for fixed P\&L, and then to try to find a desirable configuration using standard search techniques. This works in some cases, but not always. One issue is that, when alpha streams are combined on the same trading platform, some weights can be negative -- this is to be contrasted with the case when alpha streams are traded on their individual trading platforms, {\em i.e.}, a fund of funds type of a case, where the weights are non-negative. The weights being signed quantities complicates things because the weight normalization condition reads
\begin{equation}\label{weight.norm}
 \sum_{i=1}^N \left|w_i\right| = 1
\end{equation}
It is the modulus in this condition that is the source of the headache. Furthermore, when trades between different alphas are crossed, the resulting portfolio turnover reduces (compared with the case with no internal crossing) adding further complexity to the problem. In \cite{AlphaOpt} we discussed alpha stream optimization via Sharpe ratio maximization in the presence of linear costs. There, because the Sharpe ratio is invariant\footnote{\, More precisely this is the case with no costs or linear costs only.} under the rescalings $w_i\rightarrow\lambda w_i$, the condition (\ref{weight.norm}) does not affect anything, because it is ``trivially" satisfied by finding a solution without such a constraint and then simply rescaling all $w_i$ so they satisfy (\ref{weight.norm}). In the case where one does not maximize the Sharpe ratio but employs a more complex optimization criterion, scale invariance is lost and (\ref{weight.norm}) causes complications in the minimization of the objective function similar to those discussed in \cite{AlphaOpt}, but even trickier.

{}One of our main observations is that, if the alpha covariance matrix is of a factor model form, then an optimum can be found via a finite iterative procedure for a practically interesting range of the P\&L and Sharpe ratio/volatility.\footnote{\, More precisely, this is the case with no costs or linear costs, and also in the presence of nonlinear costs in a certain approximation.} In Section \ref{sec.2} we set up our notations. In Section \ref{sec.3} we discuss the case of a diagonal covariance matrix, where the aforesaid modulus issue is absent. We discuss a relaxation algorithm for finding quasi-optimal weights (which are generally expected to be close to the optimal solution when the number of alphas is large) in Subsection \ref{sub3.1}. In Subsection \ref{sub3.2} we discuss an algorithm for finding the weights when the P\&L is maximized while the Sharpe ratio is fixed. In Subsection \ref{sub3.3} we discuss a solution minimizing volatility (maximizing the Sharpe ratio) with the P\&L fixed. In Section \ref{sec4} we discuss the case of a non-diagonal covariance matrix and a trick where one effectively diagonalizes it by utilizing synthetic tradable portfolios. The problem then reduces to an iterative procedure where one needs to solve for the effective investment level. This procedure, however, generally is expected to have convergence issues which are obscured by the diagonalization trick. In Section \ref{sec5} we therefore discuss how to tackle the problem directly in the original $w_i$ basis, without diagonalizing the covariance matrix, and discuss the problem of minimizing the volatility with the P\&L fixed, which is solvable via a finite iterative procedure assuming a multi-factor form for the covariance matrix. More precisely, this is the case for the practically interesting range of the P\&L above its value corresponding to the maximum of the Sharpe ratio. We then generalize this approach to include linear costs in Section \ref{sec6} and also nonlinear costs (impact) in Section \ref{sec7}. In the latter case we employ the approximation discussed in \cite{AlphaOpt}. Section \ref{sec8} concludes the paper with some observations and comments. 

\section{Definitions}\label{sec.2}

{}We have $N$ alphas $\alpha_i$, $i=1,\dots,N$. Each alpha is actually a time series $\alpha_i(t_s)$, $s=0,1,\dots,M$, where $t_0$ is the most recent time. Below $\alpha_i$ refers to $\alpha_i(t_0)$.

{}Let $C_{ij}$ be the covariance matrix of the $N$ time series $\alpha_i(t_s)$. Let $\Psi_{ij}$ be the corresponding correlation matrix, {\em i.e.},
\begin{equation}
 C_{ij} = \sigma_i~\sigma_j~\Psi_{ij}
\end{equation}
where $\Psi_{ii} = 1$.

{}Alphas $\alpha_i$ are combined with weights $w_i$ such that
\begin{equation}\label{mod.w}
 \sum_{i=1}^N \left|w_i\right| = 1
\end{equation}
Here the modulus accounts for the fact that, when alphas are combined on the same trading platform, some weights can be negative if the covariance matrix $C_{ij}$ is not diagonal, even if all alphas are positive.

{}To begin with, we will ignore trading costs. Portfolio P\&L, volatility and Sharpe ratio are given by
\begin{eqnarray}
 &&P = I~\sum_{i=1}^N \alpha_i~w_i\\
 &&R = I~\sqrt{\sum_{i,j=1}^N C_{ij}~w_i~w_j}\\
 &&S = {P \over R}
\end{eqnarray}
where $I$ is the investment level. Our goal is to find the set of $w_i$ for which\footnote{\, In practice, one may have to restate this criterion -- see Section \ref{sec8}.}
\begin{eqnarray}\label{max}
 &&P\rightarrow{\mbox {max}}\\
 \label{Sharpe}\label{S.ineq}
 &&S \geq S_{\rm{\scriptstyle{min}}}
\end{eqnarray}
for a given Sharpe ratio lower bound $S_{\rm{\scriptstyle{min}}}$.

{}In practice, we can approach this problem in the following ways. First, we can start with the maximal possible P\&L -- which means that all weights, other than the weight corresponding to the largest alpha, are zero -- and find a relaxation by gradually reducing the P\&L until (\ref{S.ineq}) is satisfied. Second, we can solve a simplified problem $P\rightarrow{\mbox {max}}$ for a fixed value of $S = S_*$, and then use a standard searching algorithm based on a discrete set of values of $S_*$ within the desired precision. Third, we can solve another simplified problem $S\rightarrow {\mbox {max}}$ for a fixed value of $P$, and then use a standard searching algorithm based on a discrete set of values of $P$ within the desired precision. {\em I.e.}, the basic idea is to reduce the large $N$-dimensional problem to a 1-dimensional searching algorithm. In some cases this approach does work.

\section{Diagonal Case}\label{sec.3}

{}When $C_{ij}$ is not diagonal, some weights can be negative even if all alphas are positive. The modulus in (\ref{mod.w}) complicates things. However, if $C_{ij}$ is diagonal, $C_{ij} = \sigma^2_i~\delta_{ij}$, then, assuming all $\alpha_i\geq 0$ (which we can do without loss of generality), all $w_i\geq 0$ as well.\footnote{\, This is because a configuration with a negative $w_\ell$ has the same volatility as and a lower P\&L than the configuration with the sign of $w_\ell$ flipped.} Our optimization problem simplifies as follows. Let
\begin{eqnarray}\label{alphahat}
 &&{\widehat \alpha}_ i \equiv \nu_i ~\alpha_i \\
 &&{\widehat w}_i \equiv\sigma_i ~w_i\\
 &&\nu_i\equiv 1/\sigma_i
\end{eqnarray}
Then we have
\begin{eqnarray}\label{P}
 &&P = I~\sum_{i=1}^N {\widehat \alpha}_i~{\widehat w}_i\\
 &&R = I~\sqrt{\sum_{i=1}^N {\widehat w}_i^2}
\end{eqnarray}
and the optimization problem now is (\ref{max}) and (\ref{Sharpe}) subject to
\begin{eqnarray}\label{w}
 &&\sum_{i=1}^N \nu_i~{\widehat w}_i = 1\\
 &&{\widehat w}_i\geq 0
\end{eqnarray}
with all ${\widehat\alpha}_i\geq 0$. Below we discuss a simple relaxation algorithm for maximizing P\&L for $S=S_*$.

\subsection{Quasi-Optimal Weights}\label{sub3.1}

{}We can obtain a quasi-optimal solution as follows. Recall that all ${\widehat \alpha}_i$ are non-negative. First, let us label alphas such that $\alpha_1 = \max(\alpha_i, i=1,\dots,N)$. It is clear that $P$ is maximized when all ${\widehat w}_i$ vanish except for ${\widehat w}_1$. If for such ${\widehat w}_i$ the condition $S\geq S_*$ is satisfied, then this is the optimal solution. However, if it is not satisfied, then we need to sacrifice P\&L $P$ in order to increase the Sharpe ratio $S$ by allowing ${\widehat w}_2$, ${\widehat w}_3$, {\em etc.}, to be non-zero until (\ref{Sharpe}) is satisfied, where ${\widehat \alpha}_i$ are sorted as follows.\footnote{\, This sorting is a relaxation algorithm: we start with the maximal P\&L by allocating all investment into $\alpha_1$ and at each successive step add one more alpha such that P\&L is maximized when the weights are allocated to maximize the Sharpe ratio, until we reach the bound $S_*$.}

\subsubsection{Sorting Alphas}

{}For each subset
\begin{eqnarray}
 &&{\widehat w}_i > 0,~~~i=1,\dots,K\\
 &&{\widehat w}_i = 0,~~~i=K+1,\dots,N
\end{eqnarray}
the Sharpe ratio is given by (recall that ${\widehat \alpha}_i \geq 0$)
\begin{equation}
 S(K) = {{\sum_{i=1}^K {\widehat\alpha}_i~{\widehat w}_i} \over \sqrt{\sum_{i=1}^K {\widehat w}_i^2}}
\end{equation}
We want to find ${\widehat w}_i$, $i=1,\dots,K$ that maximize $S(K)$ subject to (\ref{w}):
\begin{equation}\label{w2}
 \sum_{i=1}^K \nu_i~{\widehat w}_i = 1
\end{equation}
Let us introduce a Lagrange multiplier $\mu$:
\begin{equation}
 {\widetilde S}(K) = S(K) + \mu~\left(\sum_{i=1}^K \nu_i~{\widehat w}_i - 1\right)
\end{equation}
Then the solution for ${\widehat w}_i$ that maximizes $S(K)$ subject to (\ref{w2}) is given by
\begin{eqnarray}
 &&{{\partial {\widetilde S}(K)}\over{\partial {\widehat w}_i}} = 0\\
 &&{{\partial {\widetilde S}(K)}\over{\partial \mu}} = 0
\end{eqnarray}
The $\mu$ equation gives (\ref{w2}). The ${\widehat w}_i$ equations give
\begin{eqnarray}\label{gammabeta}
 &&{\widehat w}_i = \gamma~{\widehat \alpha}_i + \beta~\nu_i,~~~i=1,\dots,K\\
 &&\gamma = {{\sum_{i=1}^K {\widehat w}^2_i}\over {\sum_{i=1}^K {\widehat \alpha}_i~{\widehat w}_i}}\label{gamma1}\\
 &&\beta = \mu~\gamma~\left(\sum_{i=1}^K {\widehat w}^2_i\right)^{1\over 2}
\end{eqnarray}
Plugging (\ref{gammabeta}) into (\ref{gamma1}), we have two solutions, $\beta = -\gamma\sum_{i=1}^K \alpha_i\nu_i / \sum_{i=1}^K \nu_i^2$, and $\beta = 0$. However, the former would imply $\sum_{i=1}^K \nu_i{\widehat w}_i = 0$, so we have the following unique solution with $\beta = 0$ and $\mu = 0$:
\begin{eqnarray}
 &&{\widehat w}_i = \gamma~{\widehat \alpha}_i\\
 &&\gamma = {1\over{\sum_{i=1}^K \nu_i~{\widehat \alpha}_i}}
\end{eqnarray}
where $\gamma$ is fixed using (\ref{w2}).

{}The corresponding P\&L in Eq. (\ref{P}) is given by
\begin{equation}
 {P(K)\over I} = {{\sum_{i=1}^K {\widehat \alpha}^2_i} \over {\sum_{i=1}^K \nu_i~{\widehat \alpha}_i}} = {{\sum_{i=1}^K \nu_i^2~\alpha^2_i} \over {\sum_{i=1}^K \nu_i^2~\alpha}_i}
\end{equation}
where we have used (\ref{alphahat}). In the remainder of this sub-subsection it will be more convenient to work with the original $\alpha_i$ as opposed to ${\widehat\alpha}_i$.

{}Let us sort $\alpha_i$ such that $P(K)$ is maximized for each $K$. {\em I.e.}, for $K=1$ we have $\alpha_1$ such that $P(1)$ is maximized (which implies $\alpha_1 = {\mbox{max}}(\alpha_i, i=1\dots,N)$), for $K=2$ we have $\alpha_1$ and $\alpha_2$, where $\alpha_1$ is fixed at $K=1$ as above, and $\alpha_2$ is such that $P(2)$ is maximized, {\em etc.} With this sorting, $P(K)$ is a monotonically decreasing function of $K$. To see this, let $\alpha_i$, $i=1,\dots,K+2$ be sorted as above ({\em i.e.}, such that at each $K$ the P\&L $P(K)$ is maximized), and let
\begin{eqnarray}
 &&a \equiv \sum_{i=1}^K \nu_i^2~\alpha_i\\
 &&b \equiv \sum_{i=1}^K \nu_i^2~\alpha^2_i\\
 &&x\equiv \alpha_{K+1}\\
 &&y\equiv \alpha_{K+2}\\
 &&u\equiv \nu^2_{K+1}\\
 &&v\equiv \nu^2_{K+2}
\end{eqnarray}
Without loss of generality, we will assume that all $\alpha_i$ are distinct, and all $\nu_i$ are distinct, $i=1,\dots,K+2$. As above, let: $P(K)$ correspond to $K$ alphas $\alpha_i$, $i=1,\dots,K$; $P(K+1)$ correspond to $K+1$ alphas $\alpha_i$, $i=1,\dots,K+1$; and $P(K+2)$ correspond to $K+2$ alphas $\alpha_i$, $i=1,\dots,K+2$. Also, let $P^\prime(K+1)$ correspond to $K+1$ alphas $\alpha^\prime_i$, $i=1,\dots,K+1$, where $\alpha^\prime_i = \alpha_i$ for $i=1,\dots,K$, and $\alpha^\prime_{K+1} = \alpha_{K+2}$. By definition, we have
\begin{equation}
 P(K+1) > P^\prime(K+1)
\end{equation}
This implies that
\begin{equation}\label{first}
 a~ u~ x^2 + v~y~\left(b + u~x^2\right) > b~ u~ x + v~ y^2~ \left(a + u~x\right)
\end{equation}
Now assume that $P(K+2) > P(K+1)$, which would imply
\begin{equation}\label{third}
 y > {{b + u~x^2}\over{a + u~x}}
\end{equation}
Together with (\ref{first}) we would then have
\begin{equation}
 x > {b\over a}
\end{equation}
and
\begin{equation}
 P(K+1) - P(K) = {{b + u~x^2}\over{a + u~x}} - {b\over a} = {{u~x~(a~x-b)}\over{a~(a + u~x)}} > 0
\end{equation}
{\em i.e.}, $P(K) < P(K+1) < P(K+2)$, and $P(K)$ must monotonically increase with $K$. However, this is not possible as $P(2) < P(1)$:
\begin{equation}
 P(2) - P(1) = \nu_2^2~\alpha_2~ {{\alpha_2 - \alpha_1}\over{\nu_1^2~\alpha_1 + \nu_2^2~\alpha}_2} < 0
\end{equation}
as $\alpha_1 = {\mbox{max}}(\alpha_i, i=1\dots,N)$. This implies that for no $K$ can $P(K+2) > P(K + 1)$, and $P(K)$ monotonically decreases with $K$.

\subsubsection{Computing Weights}

{}With the aforementioned ordering of ${\widehat \alpha}_i$, for each subset
\begin{eqnarray}
 &&{\widehat w}_i > 0,~~~i=1,\dots,K\\
 &&{\widehat w}_i = 0,~~~i=K+1,\dots,N
\end{eqnarray}
the Sharpe ratio is maximized when (see above)
\begin{equation}
 {\widehat w}_i = \gamma ~{\widehat \alpha}_i,~~~i=1,\dots,K
\end{equation}
where $\gamma>0$ is a constant. The corresponding Sharpe ratio is given by
\begin{equation}
 S(K) = \sqrt{\sum_{i=1}^K {\widehat \alpha}_i^2}
\end{equation}
The corresponding P\&L in Eq. (\ref{P}) is given by
\begin{equation}
 {P(K)\over I} = {{\sum_{i=1}^K {\widehat \alpha}^2_i} \over {\sum_{i=1}^K \nu_i~{\widehat\alpha}_i}}
\end{equation}
Here, as above, we sort ${\widehat \alpha}_i$ such that $P(K)$ is maximized for each $K$. {\em I.e.}, for $K=1$ we have ${\widehat \alpha}_1$ such that $P(1)$ is maximized, for $K=2$ we have ${\widehat \alpha}_1$ and ${\widehat \alpha}_2$, where ${\widehat \alpha}_1$ is fixed at $K=1$ as above, and ${\widehat \alpha}_2$ is such that $P(2)$ is maximized, {\em etc.} As we showed above, with this sorting, $P(K)$ is a monotonically decreasing function of $K$.

{}Let $K_*$ be such that
\begin{eqnarray}
 &&S(K_* - 1) < S_* \\
 &&S(K_*) \geq S_*
\end{eqnarray}
Then a quasi-optimal solution (see below), for which we have $S=S_*$, is given by
\begin{eqnarray}
 &&{\widehat w}_i = \gamma~{\widehat \alpha}_i,~~~i=1,\dots,K_* - 1\label{weights1}\\
 &&{\widehat w}_{K_*} = \gamma~\eta~{\widehat \alpha}_{K_*}\label{weights2}\\
 &&{\widehat w}_i = 0,~~~i=K_* + 1,\dots,N
\end{eqnarray}
where $\gamma$ will be fixed via (\ref{w}), and $0<\eta\leq 1$ is fixed by the requirement that for these weights $S = S_*$:
\begin{equation}\label{eta}
 {{\sum_{i=1}^{K_* - 1} {\widehat \alpha}_i^2 + \eta~{\widehat \alpha}_{K_*}^2} \over \sqrt{\sum_{i=1}^{K_* - 1} {\widehat \alpha}_i^2 + \eta^2~{\widehat \alpha}_{K_*}^2}} = S_*
\end{equation}
which gives\footnote{\, Eq. (\ref{eta}) has two roots, and the correct root is fixed by the requirement that $0 <\eta\leq 1$.}
\begin{equation}
 \eta = {
 {
 {\widetilde S}_1^2
 - {\widetilde S}_1~{\widetilde S}_*~
 \sqrt{ {\widetilde S}_2^2 - {\widetilde S}_*^2} }
 \over {{\widetilde S}_*^2 - 1}
 }
\end{equation}
where
\begin{eqnarray}
 &&{\widetilde S}_1 \equiv S(K_* - 1) / {\widehat \alpha_{K_*}}\\
 &&{\widetilde S}_2 \equiv S(K_*) / {\widehat \alpha_{K_*}}\\
 &&{\widetilde S}_* \equiv S_* / {\widehat \alpha_{K_*}}\\
 &&S_2^2 = S_1^2 + 1
\end{eqnarray}
Note that $\eta = 1$ if $S_* = S(K_*)$.

{}The reason why the above solution is not necessarily optimal is that $P(K)$ depends on the ordering of ${\widehat \alpha}_i$ -- we order ${\widehat \alpha}_i$ such that at each step $K\rightarrow K+1$ the P\&L $P(K+1)$ is maximized, but this does not guarantee that there is no other ordering for which the same Sharpe ratio $S_*$ (or higher) cannot be achieved for a higher P\&L, because the P\&L can be path dependent. For large $N$ this path dependence is expected to be suppressed.\footnote{\, More precisely, this is expected to be the case for generic alpha configurations.} In the next two subsections we discuss two ways of constructing the optimal solution.

\subsection{Optimal Weights}\label{sub3.2}

{}In this subsection we discuss the problem of maximizing P\&L with the Sharpe ratio fixed: $S = S_*$. We have
\begin{equation}\label{SharpeTheta}
 S = {{\sum_{i=1}^N {\widehat \alpha}_i~{\widehat w}_i} \over \sqrt{\sum_{i=1}^N {\widehat w}_i^2}}
\end{equation}
The condition $S = S_*$ can be written as
\begin{equation}\label{Theta}
 {\widehat w}^T ~\Theta ~{\widehat w} = 0
\end{equation}
where
\begin{equation}
 \Theta_{ij} \equiv {\widehat \alpha}_i~{\widehat\alpha}_j - S_*^2~\delta_{ij}
\end{equation}
The matrix $\Theta_{ij}$ has the following eigenvalue structure:
\begin{eqnarray}
 &&\Theta~\phi = \theta_*~\phi\\
 &&\Theta~\chi^{(A)} = \theta^\prime~\chi^{(A)},~~~A = 1,\dots, N-1
\end{eqnarray}
where the eigenvectors $\phi$ and $\chi^{(A)}$ and the eigenvalues $\theta_*$ and $\theta^\prime$ are given by
\begin{eqnarray}
 &&\phi_i = {{\widehat \alpha}_i\over a}\\
 &&\sum_{i=1}^N \chi^{(A)}_i~\phi_i = 0\\
 &&\theta_* = a^2  - S_*^2\\
 &&\theta^\prime = -S_*^2
\end{eqnarray}
and\footnote{\, Note that $S_{\rm{\scriptstyle{max}}} = a$, and ${\widehat w}_i = {\widehat\alpha}_i/\sum_{j=1}^N \nu_j~{\widehat\alpha}_j$ for $S=S_{\rm{\scriptstyle{max}}}$. Below we assume $S_* < S_{\rm{\scriptstyle{max}}}$.}
\begin{equation}
 a \equiv \sqrt{\sum_{i=1}^N {\widehat \alpha}_i^2}
\end{equation}
The eigenvectors $\chi^{(A)}$ are orthogonal to the eigenvector $\phi$. In fact, let us take them to be orthonormal:
\begin{equation}
 \sum_{i=1}^N \chi^{(A)}_i~\chi^{(B)}_i = \delta_{AB}
\end{equation}
where $A,B = 1,\dots, N-1$. Note that $\phi$ has norm 1 as defined above.

{}Note that the $N$ eigenvectors $\phi$ and $\chi^{(A)}$ form a complete linearly independent set of $N$-vectors. Therefore, the optimal solution ${\widehat w}_i$ can be written as
\begin{equation}
 {\widehat w}_i = \gamma~\left(\phi_i + \sum_{A=1}^{N-1} \beta_A~\chi^{(A)}_i\right)
\end{equation}
where $\gamma$ and $\beta_A$, $A=1,\dots,N-1$ are $N$ unknown coefficients we must determine. The P\&L (\ref{P}), the Sharpe ratio condition (\ref{Theta}) and the weight normalization condition (\ref{w}) now read:
\begin{eqnarray}
 &&{P \over I} = a~\gamma\label{Pgamma}\\
 &&\sum_{A=1}^{N-1} \beta_A^2 = r^2 \label{beta}\\
 &&\gamma~\left({\widetilde \phi} + \sum_{A=1}^{N-1}\beta_A~{\widetilde \chi}_A\right) = 1 \label{gamma}\\
 &&\phi_i + \sum_{A=1}^{N-1} \beta_A~\chi^{(A)}_i\geq 0,~~~i=1,\dots,N\label{wgeq0}
\end{eqnarray}
where the last condition follows from the requirement that for the optimal solution all ${\widehat w}_i\geq 0$ due to the fact that all ${\widehat\alpha}_i\geq 0$, and
\begin{eqnarray}
 &&r^2 \equiv -{\theta_*\over\theta^\prime} = {a^2\over S_*^2} - 1\\
 &&{\widetilde \phi} \equiv \sum_{i=1}^N \nu_i~\phi_i\\
 &&{\widetilde \chi}_A \equiv \sum_{i=1}^N \nu_i~\chi^{(A)}_i
\end{eqnarray}
So, our optimization problem is now reduced to maximizing $\gamma$ subject to (\ref{beta}), (\ref{gamma}) and (\ref{wgeq0}). {\em I.e.}, our optimization problem now reads:
\begin{eqnarray}\label{beta1}
 &&Y\equiv \sum_{A=1}^{N-1}\beta_A~{\widetilde \chi}_A \rightarrow \mbox{min}\\
 &&\sum_{A=1}^{N-1} \beta_A^2 = r^2\\
 &&\phi_i + \sum_{A=1}^{N-1} \beta_A~\chi^{(A)}_i\geq 0,~~~i=1,\dots,N
\end{eqnarray}
This optimization problem can be solved as follows.

{}Note that $\beta_A$ live on an $(N-2)$-sphere of radius $r$, so the sum in (\ref{beta1}) is bounded both from below and above. Also, note that without loss of generality we can always assume that ${\widetilde \chi}_A$ has the following canonical form:
\begin{eqnarray}
 &&{\widetilde \chi}_1\equiv \kappa > 0\label{chi.1}\\
 &&{\widetilde \chi}_A = 0,~~~A=2,\dots,N-1\label{chi.other}
\end{eqnarray}
We can always rotate $\chi_i^{(A)}$ and $\beta_A$ into the above canonical form using an $SO(N-1)$ rotation (under which $Y$ is invariant):
\begin{eqnarray}
 &&\chi^{(A)}_i \rightarrow \sum_{B = 1}^{N-1} U_{AB}~\chi^{(B)}_i,~~~A = 1,\dots,N-1\\
 &&{\widetilde\chi}_A \rightarrow \sum_{B = 1}^{N-1} U_{AB}~{\widetilde \chi}_B,~~~A = 1,\dots,N-1\\
 &&\beta_A \rightarrow \sum_{B = 1}^{N-1} U_{AB}~\beta_B,~~~A = 1,\dots,N-1\\
 &&\sum_{C=1}^{N-1} U_{AC}~U_{BC} = \delta_{AB},~~~A,B = 1,\dots,N-1
\end{eqnarray}
and the sign of ${\widetilde \chi}_1$ can always be set by appropriately flipping the signs of the eigenvectors $\chi^{(A)}_i$, if need be.

{}In the above basis we have
\begin{equation}
 Y = \kappa~\beta_1
\end{equation}
Its maximum occurs at $\beta_1 = r$ (the ``North Pole" on the $(N-2)$-sphere) and its minimum occurs at $\beta_1 = -r$ (the ``South Pole" on the $(N-2)$-sphere). If $\beta_1 = -r$ satisfies the constraints (\ref{wgeq0}), then this corresponds to the optimal solution. However, the constraints (\ref{wgeq0}) may not be satisfied. From the quasi-optimal solution of Subsection \ref{sub3.1} we know that a solution with ${\widehat w}_i \geq 0$ and the Sharpe ratio equal $S_*$ exists, {\em i.e.}, there exist values of $\beta_A$ on the $(N-2)$-sphere (\ref{beta}) such that (\ref{wgeq0}) are satisfied. However, this might not be the optimal solution. The optimal solution therefore lies along a path with decreasing $\beta_1$ connecting the point on the $(N-2)$-sphere corresponding to the quasi-optimal solution and the South Pole. (Note that all paths on an $(N-2)$-sphere are topologically equivalent.) Therefore, we need to identify this point corresponding to the optimal solution. One way of approaching this problem is via reducing the number of $\beta_A$ to be determined by utilizing the residual $SO(N-2)$ rotational symmetry.

{}If for the optimal solution $-r < \beta_1 < r$, then at least one $\beta_A$, $A=2,\dots,N-1$ must be non-zero. Without loss of generality we can always assume that $\beta_A$ have the following canonical form:
\begin{equation}
 \beta_A = 0,~~~A=3,\dots,N-1
\end{equation}
We can always rotate $\chi_i^{(A)}$ and $\beta_A$, $A=2,\dots,N-1$ into the above canonical form\footnote{\, Note, however, that while we are free to choose any basis for $\chi_i^{(A)}$, assuming the above canonical form implies that $\chi_i^{(2)}$ is now unknown and must be determined rather than chosen. {\em I.e.}, instead of choosing $\chi_i^{(A)}$ and trying to determine $\beta_A$, we restrict $\beta_A$ to the above canonical form at the expense of having to determine $\chi_i^{(2)}$.} using an $SO(N-2)$ rotation (under which ${\widetilde\chi}_A=0$, $A=2,\dots,N-1$ are invariant):
\begin{eqnarray}
 &&\chi^{(A)}_i \rightarrow \sum_{B = 2}^{N-1} U^\prime_{AB}~\chi^{(B)}_i,~~~A = 2,\dots,N-1\\
 &&\beta_A \rightarrow \sum_{B = 2}^{N-1} U^\prime_{AB}~\beta_B,~~~A=2,\dots,N-1\\
 &&\sum_{C=2}^{N-1} U^\prime_{AC}~U^\prime_{BC} = \delta_{AB},~~~A,B=2,\dots,N-1
\end{eqnarray}
So we have:
\begin{eqnarray}
 && \phi_i + \beta_1~\chi^{(1)}_i + \beta_2~\chi^{(2)}_i\geq 0,~~~i=1,\dots,N\\
 &&\beta_1^2 + \beta_2^2 = r^2
\end{eqnarray}
Also, since $\phi_i$ and $\chi^{(A)}_i$, $A=1,\dots,N-1$ form a complete orthonormal set of $N$-vectors, there exist coefficients $c$ and $d_A$, $A=1,\dots,N-1$ such that
\begin{equation}
 c~\phi_i + \sum_{A=1} d_A~\chi^{(A)}_i = \nu_i,~~~i=1,\dots,N
\end{equation}
which implies that
\begin{eqnarray}
 &&c = {\widetilde\phi}\\
 &&d_1 = \kappa\\
 &&d_A = 0,~~~A=2,\dots,N-1\\
 &&\chi^{(1)}_i = {1\over\kappa}~\left(\nu_i - {\widetilde\phi}~\phi_i\right)\label{chi1}\\
 &&\kappa^2 = {\widetilde\nu}^2 - {\widetilde \phi}^2
\end{eqnarray}
where
\begin{equation}
 {\widetilde\nu}^2 \equiv \sum_{i=1}^N \nu_i^2
\end{equation}
We have the following constraints:
\begin{equation}\label{chi}
 \chi^{(2)}_i \geq -{1\over{\kappa~\sqrt{r^2-\beta_1^2}}}~\left(\beta_1~\nu_i + \left[\kappa - \beta_1~{\widetilde \phi}\right]\phi_i\right)
\end{equation}
where without loss of generality we assume that $\beta_2 > 0$. (We discuss the $\beta_2 < 0$ case below.)

{}There are only four {\em a priori} conditions (which are the ${\widetilde\chi}_2 = 0$ condition plus the relevant orthonormality conditions) that the constraints (\ref{chi}) must be compatible with:
\begin{eqnarray}
 &&\sum_{i=1}^N \nu_i~\chi^{(2)}_i = 0\label{sumchi}\\
 &&\sum_{i=1}^N \phi_i~\chi^{(2)}_i = 0\label{chiphi}\\
 &&\sum_{i=1}^N \chi^{(1)}_i~\chi^{(2)}_i = 0\label{chi12}\\
 &&\sum_{i=1}^N \left[\chi^{(2)}_i\right]^2 = 1\label{chisq}
\end{eqnarray}
Due to (\ref{chi1}), (\ref{chi12}) follows from (\ref{sumchi}) and (\ref{chiphi}). Also, (\ref{chiphi}) is automatically compatible with (\ref{chi}). A nontrivial condition follows from (\ref{sumchi}) and (\ref{chi}) via multiplying both sides of (\ref{chi}) by $\nu_i$ and summing over $i=1,\dots,N$:
\begin{equation}
 {\widetilde\nu}^2 ~\beta_1 + {\widetilde\phi} \left[\kappa - \beta_1~{\widetilde \phi}\right] \geq 0
\end{equation}
which gives
\begin{equation}
 \beta_1 \geq \beta_* \equiv -{{\widetilde \phi} \over\kappa}
\end{equation}
Considering that $\beta_1 \geq -r$, this condition is satisfied if
\begin{equation}\label{phi1}
 {\widetilde \phi} \geq {\widetilde\nu}~\sqrt{1 - {S_*^2\over a^2}}
\end{equation}
If this condition is not satisfied, then $\beta_1 > -r$ and the optimal solution satisfying (\ref{wgeq0}) lies away from the South Pole. Note that (\ref{phi1}) is a necessary condition to have $\beta_1 = -r$.

{}Thus, our optimization problem has been reduced to finding the lowest value of $\beta_1$ such that there exits an $N$-vector $\chi^{(2)}$ such that (\ref{chi}), (\ref{sumchi}), (\ref{chiphi}) and (\ref{chisq}) are satisfied. Assuming $\beta_1 \geq \beta_*$, the constraints (\ref{chi}) can be incompatible with (\ref{chisq}), which further constrains $\beta_1$. For a given $\beta_1$ let $T^+$ and $T^-$ be the subsets of the values of the index $i=1,\dots,N$ such that
\begin{eqnarray}
 &&\phi_i < - {\beta_1 \over{\kappa - \beta_1~{\widetilde \phi}}}~\nu_i~,~~~i \in T^+\\
 &&\phi_i \geq - {\beta_1 \over{\kappa - \beta_1~{\widetilde \phi}}}~\nu_i~,~~~i \in T^-
\end{eqnarray}
Then we have $\chi^{(2)}_i > 0$ for $i \in T^+$ and (\ref{chisq}) may or may not be attainable for such $\beta_1$. In the following we will treat $T^+$ and $T^-$ as vectors of lengths $N^+$ and $N^-$ ($N^+ + N^- = N$), and for $i\in T^-$ we will use a map $i = T^-_a$, where $a = 1,\dots,N^-$.

{}We can construct $\chi^{(2)}$ as follows:
\begin{eqnarray}
 &&\chi^{(2)}_i = -{\theta_i\over{\kappa~\sqrt{r^2-\beta_1^2}}}~\left(\beta_1~\nu_i + \left[\kappa - \beta_1~{\widetilde \phi}\right]\phi_i\right),~~~i\in T^+\\
 &&\theta_i\geq 1,~~~i\in T^+\\
 &&\sum_{i\in T^+} \left[\chi^{(2)}_i\right]^2 \equiv {\widetilde \zeta}\\
 &&\sum_{i\in T^+} \nu_i~\chi^{(2)}_i \equiv -{\widetilde \chi}\\
 &&\sum_{i\in T^+} \chi^{(2)}_i~\phi_i \equiv -{\widetilde \xi}\\
 &&\chi^-_a \equiv \chi^{(2)}_i,~~~i=T^-_a,~~~a = 1,\dots,N^-\\
 &&\varphi_a \equiv \phi_i,~~~i=T^-_a,~~~a = 1,\dots,N^-\\
 &&\omega_a \equiv \nu_i,~~~i=T^-_a,~~~a = 1,\dots,N^-
\end{eqnarray}
The $N^-$-vector $\chi^-_a$ is constrained as follows:
\begin{eqnarray}
 &&\sum_{a=1}^{N^-} \omega_a~\chi^-_a = {\widetilde \chi}\\
 &&\sum_{a=1}^{N^-} \chi^-_a~\varphi_a = {\widetilde \xi}\\
 &&\sum_{a=1}^{N^-} \left[\chi^-_a\right]^2 = 1 - {\widetilde \zeta}\label{chisq1}
\end{eqnarray}
The vector $\chi^-_a$ can always be decomposed as follows:
\begin{eqnarray}
 &&\chi^-_a = c_- ~ \varphi_a + d_-~\omega_a + {\widehat \chi}^-_a\\
 &&c_- \equiv {{{\widetilde\lambda}~{\widetilde \xi} - {\widetilde \varphi}~{\widetilde \chi}}\over{{\widetilde \sigma}~{\widetilde\lambda} - {\widetilde \varphi}^2}}\\
 &&d_- \equiv {{{\widetilde \sigma}~{\widetilde \chi} - {\widetilde \varphi}~{\widetilde \xi}}\over{{\widetilde \sigma}~{\widetilde\lambda} - {\widetilde \varphi}^2}}
\end{eqnarray}
where
\begin{eqnarray}
 &&{\widetilde \sigma} \equiv \sum_{a=1}^{N^-} \varphi_a^2\\
 &&{\widetilde\varphi} \equiv \sum_{a=1}^{N^-} \omega_a~\varphi_a\\
 &&{\widetilde\lambda} \equiv \sum_{a=1}^{N^-} \omega_a^2
\end{eqnarray}
and the $N^-$-vector ${\widehat \chi}^-_a$ is such that
\begin{eqnarray}
 &&\sum_{a=1}^{N^-} {\widehat \chi}^-_a~\omega_a = 0\\
 &&\sum_{a=1}^{N^-} {\widehat \chi}^-_a~\varphi_a = 0
\end{eqnarray}
With this decomposition, (\ref{chisq1}) reads
\begin{eqnarray}\label{chisq2}
 &&\sum_{a=1}^{N^-} \left[{\widehat \chi}^-_a\right]^2 = 1 - {\widetilde \zeta} - {g({\widetilde \chi},{\widetilde\xi})\over{{\widetilde\sigma}~{\widetilde\lambda} - {\widetilde\varphi}^2}}\\
 &&g(x_1,x_2)\equiv {\widetilde\sigma}~x_1^2 + {\widetilde\lambda}~x_2^2 - 2~{\widetilde\varphi}~x_1~x_2
\end{eqnarray}
Note that
\begin{eqnarray}
 &&{\widetilde \sigma} = \sum_{i\in T^-} \phi_i^2 = {{\sum_{i\in T^-} {\widehat\alpha}_i^2}\over {\sum_{i=1}^N {\widehat\alpha}_i^2}} \leq 1\\
 &&{{\widetilde\varphi}^2 \over{\widetilde\sigma}} = {\left(\sum_{i\in T^-} \nu_i~{\widehat\alpha}_i\right)^2 \over {\sum_{i\in T^-} {\widehat\alpha}_i^2}} < {\widetilde \lambda} = \sum_{i\in T^-} \nu_i^2
\end{eqnarray}
where we assume that ${\widehat \alpha}_i$, $i\in T^-$ are not all identical, and $\nu_i$, $i\in T^-$ are not all identical. Further, note that each of ${\widetilde\zeta}$, ${\widetilde \xi}$ and ${\widetilde\chi}$ is minimized when $\theta_i\equiv 1$, $i\in T^+$, while if we fix $\theta_i$, then ${\widetilde\zeta}$ increases as $\beta_1$ decreases (assuming $\beta_1 <0$). Finally, the eigenvalues of the $2\times 2$ curvature matrix
\begin{equation}
 {{\partial^2 g(x_1, x_2)}\over{\partial x_p~\partial x_q}},~~~p,q = 1,2
\end{equation}
are given by
\begin{equation}
 g_\pm ={1\over 2}~\left({\widetilde \lambda} + {\widetilde \sigma} \pm \sqrt{\left({\widetilde\lambda} - {\widetilde \sigma}\right)^2 + 4~{\widetilde\varphi}^2}\right)
\end{equation}
and are both positive considering that ${\widetilde\varphi}^2 < {\widetilde\sigma}~{\widetilde\lambda}$. This implies that the r.h.s. of (\ref{chisq2}) is maximized when $\theta_i\equiv 1$, $i\in T^+$. Therefore, the above construction provides $\chi^{(2)}_i$ corresponding to the optimal ({\em i.e.}, minimal) $\beta_1$ when $\theta_i\equiv 1$, $i\in T^+$ and ${\widehat \chi}^-_a\equiv 0$, $a=1,\dots,N^-$:
\begin{eqnarray}
 &&\chi^{(2)}_i = -{1\over{\kappa~\sqrt{r^2-\beta_1^2}}}~\left(\beta_1~\nu_i + \left[\kappa - \beta_1~{\widetilde \phi}\right]\phi_i\right),~~~i\in T^+\\
 &&\chi^{(2)}_i = c_- ~ \phi_i + d_-~\nu_i,~~~i\in T^-\\
 &&{\widetilde \zeta} + {g({\widetilde \chi},{\widetilde\xi})\over{{\widetilde\sigma}~{\widetilde\lambda} - {\widetilde\varphi}^2}} = 1\label{optbeta}
\end{eqnarray}
The last equation (\ref{optbeta}) is then used to determine the optimal value of $\beta_1$. Practically, one needs to employ an iterative procedure to determine $\beta_1$ from (\ref{optbeta}) because $T^+$ and $T^-$ depend on $\beta_1$. One way of implementing such an iterative procedure is to use the quasi-optimal solution of the previous subsection. Let $z_i\equiv {\widehat w}_i$ for that solution. Then the corresponding value of $\beta_1\equiv {\widehat \beta}_1$ is given by (this follows from (\ref{Pgamma}), (\ref{gamma}), (\ref{chi.1}) and (\ref{chi.other}))
\begin{equation}
 {\widehat\beta}_1 = {{1 -{\widetilde\phi}~\sum_{i=1}^N z_i~\phi_i}\over{\kappa~\sum_{i=1}^N z_i~\phi_i}}
\end{equation}
We can now search for the value of $\beta_1$ between ${\widehat\beta}_1$ and $\beta_*$, {\em e.g.}, using a standard algorithm such as successively testing midpoints until the l.h.s. of (\ref{optbeta}) approaches 1 from below with a desired precision.

{}Above we assume that $\beta_2 > 0$. If we assume that $\beta_2 <0$, then instead of (\ref{chi}) we have:
\begin{equation}
 \chi^{(2)}_i \leq {1\over{\kappa~\sqrt{r^2-\beta_1^2}}}~\left(\beta_1~\nu_i + \left[\kappa - \beta_1~{\widetilde \phi}\right]\phi_i\right)
\end{equation}
and the above discussion carries through unchanged except that now $\chi^{(2)}_i$, $i\in T^+$ are all negative.

{}Finally, the weights ${\widehat w}_i = 0$ for $i\in T^+$, while for $i\in T^-$ they are given by
\begin{equation}
 {\widehat w}_i = {{\left(1 - {\beta_1\over \kappa}~{\widetilde \phi} + \sqrt{r^2 - \beta_1^2}~c_-\right)\phi_i + \left({\beta_1\over\kappa} + \sqrt{r^2 - \beta_1^2}~d_-\right)\nu_i}
 \over{{\widetilde\phi} + \beta_1~\kappa}}
\end{equation}
The fact that non-zero ${\widehat w}_i$ have a form ${\widehat w}_i = e~\phi_i + f~\nu_i$ (where $e$ and $f$ are coefficients) is not surprising from symmetry considerations: $\phi_i$ and $\nu_i$ are the only vectors available as building blocks. However, what is nontrivial is the definition of $T^+$ and $T^-$, {\em i.e.}, which ${\widehat w}_i$ are vanishing.

\subsection{Alternative Construction}\label{sub3.3}

{}We can construct the above optimal solution in a different way. Instead of maximizing P\&L directly, we can i) first minimize the volatility $R$ ({\em i.e.}, maximize the Sharpe ratio $S$) for a fixed value of the P\&L $P$, and ii) then find $P$ such that $S=S_*$. {\em I.e.}, at the first step we solve the following problem:
\begin{eqnarray}
 &&{P \over I} = \sum_{i=1}^N {\widehat \alpha}_i~{\widehat w}_i \equiv {\widetilde P} = \mbox{fixed}\\
 &&{R \over I} = \sqrt{\sum_{i=1}^N {\widehat w}_i^2} \rightarrow \mbox{min}\\
 &&\sum_{i=1}^N \nu_i~{\widehat w}_i = 1\\
 &&{\widehat w}_i \geq 0,~~~i=1,\dots,N
\end{eqnarray}
This problem can be stated as follows:
\begin{eqnarray}
 &&g({\widehat w}, \mu, {\widetilde \mu}) \equiv {1\over 2}~\sum_{i=1}^N {\widehat w}_i^2 + \mu~\left(\sum_{i=1}^N \nu_i~{\widehat w}_i - 1\right) +
 {\widetilde \mu}~\left(\sum_{i=1}^N {\widehat \alpha}_i~{\widehat w}_i - {\widetilde P}\right)\\
 &&g({\widehat w}, \mu, {\widetilde \mu}) \rightarrow \mbox{min}\\
 &&{\widehat w}_i \geq 0,~~~i=1,\dots,N
\end{eqnarray}
where $\mu$ and ${\widetilde \mu}$ are Lagrange multipliers, {\em i.e.}, the objective function $g({\widehat w}, \mu, {\widetilde \mu})$ is minimized w.r.t. ${\widehat w}_i$, $\mu$ and ${\widetilde \mu}$.

{}The solution is as follows. First, let us sort $\alpha_i$ ({\em not} ${\widehat \alpha}_i$) in the decreasing order (recall that ${\widehat \alpha}_i = \nu_i~\alpha_i$), {\em i.e.}, $\alpha_1 = \mbox{max}(\alpha_i, i=1,\dots,N)$. Then
\begin{eqnarray}
 &&{\widehat w}_i = {{\left({\widetilde\nu}^2~{\widetilde P} - a\right){\widehat \alpha}_i - \left(a~{\widetilde P} - b^2\right)\nu_i}\over{{\widetilde\nu}^2~b^2 - a^2}},~~~i=1,\dots,K\\
 &&{\widehat w}_i = 0,~~~i=K+1,\dots,N\\
 &&a \equiv \sum_{i=1}^K \nu_i~{\widehat \alpha}_i\\
 &&b^2 \equiv \sum_{i=1}^K {\widehat \alpha}^2_i\\
 &&{\widetilde\nu}^2\equiv \sum_{i=1}^K \nu_i^2\\
 &&{\widehat \alpha}_i > {{a~{\widetilde P} - b^2}\over {{\widetilde\nu}^2~{\widetilde P} - a}}~\nu_i,~~~i=1,\dots,K\label{K-cond}
\end{eqnarray}
where the last condition determines $K$, {\em i.e.}, $K$ is the maximum number of first $K$ alphas $\alpha_i$ (sorted in the decreasing order) such that (\ref{K-cond}), which is equivalent to
\begin{equation}
 \alpha_i > {{a~{\widetilde P} - b^2}\over {{\widetilde\nu}^2~{\widetilde P} - a}},~~~i=1,\dots,K
\end{equation}
is satisfied. Note that ${\widetilde\nu}^2~b^2 > a^2$ (assuming non-identical ${\widehat \alpha}_i$ and non-identical $\nu_i$). Also, ${\widetilde P}\leq \alpha_1$, and we can assume ${\widetilde P}>0$. Finally, note that ${\widetilde P} = b^2/a$ corresponds to
\begin{equation}\label{S.max.1}
 S = S_{\rm{\scriptstyle{max}}} \equiv \sqrt{\sum_{i=1}^N {\widehat\alpha}_i^2}
\end{equation}
Indeed, in this case we have ${\widehat w}_i = {\widehat\alpha}_i / a$ (and all ${\widehat w}_i$ are non-vanishing unless some ${\widehat \alpha}_i = 0$). Therefore, we can assume that ${\widetilde P}\geq b^2/a$, because, if we consider lower ${\widetilde P}$, we have a solution with higher Sharpe ratio and higher P\&L, so the search can be limited to the values $b^2/a < {\widetilde P} < \alpha_1$.

{}Using the above algorithm we can construct ${\widehat w}_i$ for any given ${\widetilde P}$. We can then determine the weights corresponding to $S=S_*$ using a standard algorithm such as successively testing midpoints between successive values of ${\widetilde P}$ (between $\alpha_1$ and $b^2/a$) until the Sharpe ratio $S$ approaches $S_*$ with a desired precision.

\section{Non-Diagonal Case}\label{sec4}

{}When the covariance matrix $C_{ij}$ is non-diagonal, things are more complicated because of the modulus in (\ref{mod.w}). We can still go into a diagonal basis via the following trick. Let $V_i^{(a)}$ be $N$ right eigenvectors of $C_{ij}$ corresponding to its eigenvalues $\lambda^{(a)}$, $a=1,\dots,N$:
\begin{equation}
 C~V^{(a)} = \lambda^{(a)}~V^{(a)}
\end{equation}
with no summation over $a$. Since $C$ is symmetric, $V^{(a)}$ can be chosen to be orthonormal:
\begin{equation}
 \sum_{i=1}^N V^{(a)}_i~V^{(b)}_i = \delta_{ab}
\end{equation}
which we assume to be the case. We will also assume that all $\lambda^{(a)} > 0$, {\em i.e.}, $C_{ij}$ is positive-definite.\footnote{\, A simple method for making a covariance matrix positive-definite was discussed in \cite{SpMod} based on \cite{RJ}.} Furthermore, we will assume that none of the eigenvalues $\lambda^{(a)}$ is ``infinitesimally" small compared with others.\footnote{\, {\em I.e.}, none of these eigenvalues are zeros distorted by machine precision, as would be the case if $M < N$ (where $M+1$ is the number of observations in the alpha time series -- see Section \ref{sec.2}).}

{}Now consider $N$ synthetic portfolios $\Pi^{(a)}$, $a=1,\dots,N$, with the weights
\begin{equation}
 w^{(a)}_i = {V_i^{(a)} \over {\sum_{j=1}^N \left|V_j^{(a)}\right|}}
\end{equation}
Note that
\begin{equation}
 \sum_{i=1}^N \left|w^{(a)}_i\right| = 1
\end{equation}
Since we are assuming that all alphas are traded on the same execution platform, the portfolios $\Pi^{(a)}$ are tradable. Furthermore, the covariance matrix for these portfolios is diagonal:
\begin{equation}
 C_{ab}\equiv \left\langle \alpha^{(a)}, \alpha^{(b)}\right\rangle = {\lambda^{(a)}\over\left(\sum_{j=1}^N \left|V_j^{(a)}\right|\right)^2}~\delta_{ab}
\end{equation}
The corresponding alphas are given by
\begin{equation}
 \alpha^{(a)}\equiv {1\over {\sum_{j=1}^N\left|V_j^{(a)}\right|}} \sum_{i = 1}^N V_i^{(a)}~\alpha_i
\end{equation}
and can be assumed to be non-negative -- if any $\alpha^{(a)}$ is negative, we can always make it positive by flipping the signs of the weights in the corresponding synthetic portfolio. Now we could proceed to solve the optimization problem for these synthetic alphas as we did in the previous section. However, the caveat is that if we invest $I$ dollars into a combinations of these synthetic $\alpha^{(a)}$, it does not correspond to investing $I$ dollars into the underlying alphas due to ``netting" of alphas\footnote{\, Not to be confused with the ``netting" of underlying tradables.} as the weights $w^{(a)}_i$ in various portfolios have opposite signs. So, in the optimization problem, on paper, we need to invest some {\em priori} unknown ``synthetic" amount $I^\prime$ into the synthetic alphas $\alpha^{(a)}$ to have the desired actual investment $I$ in the real alphas $\alpha_i$. This amount $I^\prime$ would then have to be determined via an iterative procedure. However, for general $C_{ij}$ this iterative procedure can run into stability issues. This is because the optimization now becomes effectively 2-dimensional. In this regard, it is more streamlined to tackle the problem by dealing with the modulus in (\ref{mod.w}) directly as opposed to attempting to circumvent it via diagonalization. We discuss this approach in the next section.

\section{Optimal Weights}\label{sec5}

{}Here we will follow the framework of Subsection \ref{sub3.3} and i) first minimize the volatility $R$ ({\em i.e.}, maximize the Sharpe ratio $S$) for a fixed value of the P\&L $P$, and ii) then find $P$ such that $S=S_*$. The second step is straightforward.\footnote{\, The second step is a one-dimensional root searching problem.} At the first step we solve the following problem:
\begin{eqnarray}
 &&{P \over I} = \sum_{i=1}^N \alpha_i~w_i \equiv {\widetilde P} = \mbox{fixed}\\
 &&{R \over I} = \sqrt{\sum_{i,j=1}^N C_{ij}~w_i~w_j} \equiv {\widetilde R} \rightarrow \mbox{min}
\end{eqnarray}
subject to 
\begin{equation}
 \sum_{i=1}^N |w_i| = 1
\end{equation}
where $C_{ij}$ is arbitrary except that, without loss of generality for our purposes here (see below), we will assume it to be positive-definite. This problem can be stated as follows:
\begin{eqnarray}
 &&g(w, \mu, {\widetilde \mu}) \equiv {1\over 2}~\sum_{i,j=1}^N C_{ij}~w_i~w_j + \nonumber \\
 &&~~~ + \mu\left(\sum_{i=1}^N \left|w_i\right| - 1\right) +
 {\widetilde \mu}\left(\sum_{i=1}^N \alpha_i~w_i - {\widetilde P}\right)\rightarrow \mbox{min}\label{g}
\end{eqnarray}
where $\mu$ and ${\widetilde \mu}$ are Lagrange multipliers, {\em i.e.}, the objective function $g(w, \mu, {\widetilde \mu})$ is minimized w.r.t. $w_i$, $\mu$ and ${\widetilde \mu}$. What complicates matters is the modulus of $w_i$. The problem can still be solved, albeit it requires a finite iterative procedure, {\em i.e.}, the solution is exact and is obtained after a finite number of iterations.\footnote{\, More precisely, the optimum is unique for $\mu \geq 0$ -- see below.}

{}Let $J$ and $J^\prime$ be the subsets of the index $i=1,\dots,N$ such that
\begin{eqnarray}
 &&w_i\not= 0,~~~i\in J\\
 &&w_i = 0,~~~i\in J^\prime
\end{eqnarray}
Let
\begin{equation}\label{eta.i}
 \eta_i \equiv \mbox{sign}\left(w_i\right),~~~i\in J
\end{equation}
Note that, since the modulus has a discontinuous derivative, the minimization equations are not the same as setting first derivatives of $g(w, \mu, {\widetilde \mu})$ to zero. More concretely, first derivatives are well-defined for $i\in J$, but not for $i\in J^\prime$. So, we have the following minimization equations for $w_i$, $i\in J$:
\begin{eqnarray}\label{J1}
 &&\sum_{j\in J} C_{ij}~w_j + \mu~\eta_i + {\widetilde\mu} ~\alpha_i = 0,~~~i\in J\\
 &&\sum_{i\in J} \left|w_i\right| = 1\label{J2}\\
 &&\sum_{i\in J} \alpha_i~w_i = {\widetilde P}\label{J3}
\end{eqnarray}
with $\mu$ and ${\widetilde \mu}$ determined using (\ref{J2}) and (\ref{J3}).
There are additional conditions for the global minimum corresponding to the directions $i\in J^\prime$:
\begin{eqnarray}\label{JJ1}
 &&\sum_{i,j=1}^N C_{ij}~\left(w_i + \epsilon_i\right)~\left(w_j + \epsilon_j\right) \geq \sum_{i,j \in J} C_{ij}~w_i~w_j\\
 &&\sum_{i=1}^N \left|w_i + \epsilon_i\right| = 1\label{JJ2}\\
 &&\sum_{i=1}^N \alpha_i~\left(w_i + \epsilon_i\right) = {\widetilde P}\label{JJ3}
\end{eqnarray}
where $w_i$, $i\in J$ are determined using (\ref{J1}), while $w_i =0$, $i\in J^\prime$. The conditions (\ref{JJ1}) must be satisfied including for arbitrary infinitesimal $\epsilon_i$ subject to (\ref{JJ2}) and (\ref{JJ3}). For infinitesimal $\epsilon_i$ these conditions can be rewritten as follows:\footnote{\, Since here $\epsilon_i$ are taken to be infinitesimal, these are the conditions for a local minimum. See Subsection \ref{sub5.5} for the global minimum conditions.}
\begin{eqnarray}
 &&\sum_{i,j \in J} C_{ij}~w_i~\epsilon_j + \sum_{i\in J} \sum_{j\in J^\prime} C_{ij}~w_i~\epsilon_j \geq 0\\
 &&\sum_{i\in J} \eta_i~\epsilon_i + \sum_{i\in J^\prime} \left|\epsilon_i\right| = 0\\
 &&\sum_{i\in J} \alpha_i~\epsilon_i + \sum_{i\in J^\prime} \alpha_i~\epsilon_i = 0
\end{eqnarray}
which, taking into account (\ref{J1}), reduce to
\begin{equation}\label{JJ4}
 \sum_{j\in J^\prime}\left(\sum_{i\in J} C_{ij}~w_i~\epsilon_j + \mu~\left|\epsilon_j\right| + {\widetilde \mu}~\alpha_j~\epsilon_j\right)\geq 0
\end{equation}
Since $\epsilon_j$, $j\in J^\prime$ are arbitrary, this gives the following conditions:
\begin{equation}\label{globalmin}
 \forall j\in J^\prime:~~~\left|{\widetilde\mu}~\alpha_j + \sum_{i\in J} C_{ij}~w_i\right| \leq \mu
\end{equation}
These conditions must be satisfied by the solution to (\ref{J1}), (\ref{J2}) and (\ref{J3}). The solution that (locally) minimizes $g(w, \mu, {\widetilde \mu})$ is given by (in matrix notation)
\begin{eqnarray}\label{w-eta}
 &&w = -\mu~D~\eta -{\widetilde \mu} ~D~\alpha\\
 &&\mu = {{{\widetilde P}~(\alpha^T~D~\eta) - (\alpha^T~D~\alpha)}\over{(\alpha^T~D~\alpha)~(\eta^T~D~\eta) - (\alpha^T~D~\eta)^2}}\\
 &&{\widetilde\mu} = {{(\alpha^T~D~\eta) - {\widetilde P}~(\eta^T~D~\eta)}\over{(\alpha^T~D~\alpha)~(\eta^T~D~\eta) - (\alpha^T~D~\eta)^2}}
\end{eqnarray}
Here all (both contracted and free) indices are assumed to run over $i\in J$, {\em e.g.},
\begin{equation}\label{w.sol.gen}
 w_i = -\mu~\sum_{j\in J} D_{ij}~\eta_j - {\widetilde \mu} ~\sum_{j\in J} D_{ij}~\alpha_j,~~~i\in J
\end{equation}
and $D$ is the inverse matrix of the $N(J)\times N(J)$ matrix $C_{ij}$, $i,j\in J$, where $N(J)\equiv\left|J\right|$ is the number of elements of $J$:
\begin{equation}\label{D}
 \sum_{k\in J} C_{ik}~D_{kj} = \delta_{ij},~~~i,j\in J
\end{equation}
{\em i.e.}, $D$ is {\em not} a restriction of the inverse of the $N\times N$ matrix $C_{ij}$, $i,j\in 1,\dots,N$ to $i,j\in J$.

{}Here the following observation is in order. In the above solution, {\em a priori} we do not know i) what the subset $J^\prime$ is and ii) what the values of $\eta_i$ are for $i\in J$. This means that {\em a priori} we have total of $3^N$ possible combinations (including the redundant empty $J$ case), so if we go through this finite set, we will solve the problem exactly. However, $3^N$ is a prohibitively large number,\footnote{\, We assume that $N$ itself is large.} so one needs a more clever way of solving the problem.\footnote{\, We need to iterate because of the term proportional to $\mu$ in (\ref{w.sol.gen}). Let us assume $\mu = 0$. Then for generic alpha configurations $J^\prime$ is empty, $D = C^{-1}$, and the solution is $w = (\alpha^T C^{-1} \eta)^{-1}C^{-1}\alpha$ with ${\widetilde P} = (\alpha^T C^{-1}\alpha)/(\alpha^T C^{-1} \eta)$, which corresponds to the solution to $S\rightarrow \mbox{max}$ with $S=S_{\rm{\scriptstyle{max}}}=\sqrt{\alpha^T C^{-1}\alpha}$. We discuss this case in more detail in Subsection \ref{sub5.3} in the context of the factor model covariance matrix.}

\subsection{Factor Model}\label{sub5.1}
{}As was pointed out in \cite{AlphaOpt} in the context of optimization with linear costs, a way to reduce the number of iterations in this types of problems is to reduce the ``off-diagonality" of $C_{ij}$. This can be achieved by considering a factor model for alphas.\footnote{\, Here we simply assume a factor model form for the covariance matrix without delving into details of how it is constructed.}

{}As in the case of a stock multi-factor risk model, instead of $N$ alphas, one deals with $F\ll N$ risk factors and the covariance matrix $C_{ij}$ is replaced by $\Gamma_{ij}$ given by
\begin{eqnarray}\label{Gamma}
 &&\Gamma \equiv \Xi + \Omega~\Phi~\Omega^T \equiv \Xi + {\widetilde \Omega}~{\widetilde \Omega}^T\\
 && \Xi_{ij} \equiv \xi_i^2 ~\delta_{ij}\\
 && {\widetilde \Phi}~{\widetilde \Phi}^T = \Phi
\end{eqnarray}
where $\xi_i$ is the specific risk for each $\alpha_i$; $\Omega_{iA}$ is an $N\times F$ factor loadings matrix; ${\widetilde \Omega}\equiv \Omega~{\widetilde\Phi}$; $\Phi_{AB}$ is the factor covariance matrix, $A,B=1,\dots,F$; and ${\widetilde \Phi}_{AB}$ is the Cholesky decomposition of $\Phi_{AB}$, which is assumed to be positive-definite. {\em I.e.}, the random processes $\Upsilon_i$ corresponding to $N$ alphas are modeled via $N$ random processes $z_i$ (corresponding to specific risk) together with $F$ random processes $f_A$ (corresponding to factor risk):
\begin{eqnarray}
 &&\Upsilon_i = z_i + \sum_{A=1}^F \Omega_{iA}~f_A\\
 &&\left<z_i, z_j\right> = \Xi_{ij}\\
 &&\left<z_i, f_A\right> = 0\\
 &&\left<f_A, f_B\right> = \Phi_{AB}\\
 &&\left<\Upsilon_i, \Upsilon_j\right> = \Gamma_{ij}
\end{eqnarray}
Instead of an $N \times N$ covariance matrix $C_{ij}$ we now have an $F \times F$ covariance matrix $\Phi_{AB}$. In the following we assume that $C_{ij}$ has the factor model form $\Gamma_{ij}$.

\subsection{Optimization with Factor Model}\label{sub5.2}

{}In this framework, the problem reduces to solving an $(F+2)$-dimensional system as follows. First, let
\begin{equation}\label{v}
 v_A \equiv \sum_{i=1}^N~w_i~{\widetilde \Omega}_{iA} = \sum_{i\in J}~w_i~{\widetilde \Omega}_{iA},~~~A=1,\dots,F
\end{equation}
Then from (\ref{J1}) we have
\begin{equation}\label{wv}
 w_i = - {1\over \xi_i^2}~\left(\mu~\eta_i + {\widetilde \mu}~\alpha_i + \sum_{A=1}^F {\widetilde \Omega}_{iA}~v_A\right),~~~i\in J
\end{equation}
Recalling that $w_i~\eta_i > 0$, $i\in J$, and {\em assuming} that $\mu \geq 0$ (see below), we get
\begin{eqnarray}
 &&\eta_i = -\mbox{sign}\left({\widetilde\mu}~\alpha_i + \sum_{A=1}^F {\widetilde \Omega}_{iA}~v_A\right),~~~i\in J\label{eta1}\\
 &&\forall i\in J:~~~\left|{\widetilde\mu}~\alpha_i + \sum_{A=1}^F {\widetilde \Omega}_{iA}~v_A\right| > \mu\label{eta2}\\
 &&\forall i\in J^\prime:~~~\left|{\widetilde\mu}~\alpha_i + \sum_{A=1}^F {\widetilde \Omega}_{iA}~v_A\right| \leq \mu\label{eta3}
\end{eqnarray}
where (\ref{eta2}) follows from (\ref{wv}), and (\ref{eta3}) follows from (\ref{globalmin}). The last two inequalities define $J$ and $J^\prime$ in terms of $(F+2)$ unknowns $v_A$, $\mu$ and ${\widetilde \mu}$.

{}In the remainder of this subsection we assume that $\mu\geq 0$. We discuss the $\mu < 0$ case in Subsection \ref{sub5.4}.

{}Substituting (\ref{wv}) into (\ref{v}), (\ref{J2}) and (\ref{J3}), we get the following system of $(F+2)$ equations for $(F+2)$ unknowns $v_A$, $\mu$ and ${\widetilde \mu}$:
\begin{eqnarray}
 &&\sum_{B=1}^F Q_{AB}~v_B + a_A~\mu + b_A~{\widetilde \mu} = 0\\
 &&\sum_{A=1}^F a_A~v_A + c~\mu + d~{\widetilde \mu} = -1\\
 &&\sum_{A=1}^F b_A~v_A + d~\mu + e~{\widetilde \mu} = -{\widetilde P}
\end{eqnarray}
where
\begin{eqnarray}
 &&Q_{AB} \equiv \delta_{AB} + \sum_{i\in J} {{{\widetilde \Omega}_{iA}~{\widetilde \Omega}_{iB}} \over {\xi_i^2}}\\
 &&a_A \equiv \sum_{i\in J} {{\eta_i~{\widetilde \Omega}_{iA}} \over {\xi_i^2}}\\
 &&b_A \equiv \sum_{i\in J} {{\alpha_i~{\widetilde \Omega}_{iA}} \over {\xi_i^2}}\\
 &&c \equiv \sum_{i\in J} {1 \over {\xi_i^2}}\\
 &&d \equiv \sum_{i\in J} {{\alpha_i~\eta_i} \over {\xi_i^2}}\\
 &&e \equiv \sum_{i\in J} {{\alpha_i^2} \over {\xi_i^2}}
\end{eqnarray}
Let ${\widetilde Q}_{ab}$, $a,b = 1,\dots,(F+2)$ be the following symmetric $(F+2) \times (F+2)$ matrix:
\begin{eqnarray}
 &&{\widetilde Q}_{AB} \equiv Q_{AB}\\
 &&{\widetilde Q}_{A,(F+1)} = {\widetilde Q}_{(F+1), A} \equiv a_A\\
 &&{\widetilde Q}_{A,(F+2)} = {\widetilde Q}_{(F+2), A} \equiv b_A\\
 &&{\widetilde Q}_{(F+1),(F+1)} \equiv c\\
 &&{\widetilde Q}_{(F+1),(F+2)} = {\widetilde Q}_{(F+2),(F+1)}\equiv d\\
 &&{\widetilde Q}_{(F+2),(F+2)} \equiv e
\end{eqnarray}
Also, let $x_a$ and $y_a$, $a=1,\dots,(F+2)$ be the following $(F+2)$-vectors:
\begin{eqnarray}
 &&x_A \equiv v_A\\
 &&x_{F+1} \equiv \mu\\
 &&x_{F+2} \equiv {\widetilde \mu}\\
 &&y_A = 0\\
 &&y_{F+1} = -1\\
 &&y_{F+2} = -{\widetilde P}
\end{eqnarray}
Then we have
\begin{equation}\label{F+2}
 x_a = \sum_{b=1}^{F+2} {\widehat Q}_{ab}~y_b
\end{equation}
where ${\widehat Q}_{ab}$ is the matrix inverse to ${\widetilde Q}_{ab}$: ${\widehat Q} \equiv {\widetilde Q}^{-1}$.

{}Note that (\ref{F+2}) solves for $v_A$, $\mu$ and ${\widetilde \mu}$ given $\eta_i$, $J$ and $J^\prime$. On the other hand, (\ref{eta1}), (\ref{eta2}) and (\ref{eta3}) determine $\eta_i$, $J$ and $J^\prime$ in terms of $v_A$, $\mu$ and ${\widetilde \mu}$. The entire system is then solved iteratively, where at the initial iteration one takes $J^{(0)}=\{1,\dots,N\}$, so that $J^{\prime(0)}$ is empty, and
\begin{equation}
 \eta^{(0)}_i = \pm 1,~~~i=1,\dots,N
\end{equation}
While {\em a priori} the values of $\eta^{(0)}_i$ can be arbitrary, unless $F\ll N$, in some cases one might encounter convergence issues. However, if one chooses
\begin{equation}
 \eta^{(0)}_i = \mbox{sign}(\alpha_i),~~~i=1,\dots,N
\end{equation}
then the iterative procedure generally is expected to converge rather fast. Furthermore, note that the solution is actually exact, {\em i.e.}, the convergence criteria are given by (from Subsection \ref{sub5.5} we have that for $\mu \geq 0$ this produces the global optimum)
\begin{eqnarray}
 &&J^{(s + 1)} = J^{(s)}\\
 &&\forall i\in J^{(s+1)}:~~~\eta^{(s+1)}_i = \eta^{(s)}_i\\
 &&\mu^{(s + 1)} = \mu^{(s)}\\
 &&{\widetilde \mu}^{(s + 1)} = {\widetilde \mu}^{(s)}\\
 &&\forall A\in \{1,\dots,F\}:~~~v^{(s + 1)}_A = v^{(s)}_A
\end{eqnarray}
where $s$ and $s+1$ label successive iterations.\footnote{\, The first two of these criteria are based on discrete quantities and are unaffected by computational (machine) precision effects, while the last three of these criteria are based on continuous quantities and in practice are understood as satisfied within computational (machine) precision.
In practice, it suffices to check only one of the three continuous criteria, {\em e.g.}, the convergence of $\mu$.} Put differently, the iterative procedure is finite -- it converges in a finite number of iterations. Finally, note that $w_i$ for $i\in J$ are given by (\ref{wv}), while $w_i = 0$ for $i\in J^\prime$.

\subsection{$\mu = 0$ Case}\label{sub5.3}

{}From the previous subsection we have
\begin{eqnarray}
 &&\mu = {{{\widetilde d}~{\widetilde P} - {\widetilde e}}\over{{\widetilde c}~{\widetilde e} - {\widetilde d}^2}}\label{mu.2}\\
 &&{\widetilde \mu} = {{{\widetilde d} - {\widetilde c}~{\widetilde P}}\over{{\widetilde c}~{\widetilde e} - {\widetilde d}^2}}\label{mutilde.2}\\
 &&0 < {\widetilde R}^2 = 2\left.g(w_i,\mu,{\widetilde\mu})\right|_{\rm{\scriptstyle{optimum}}} = -\mu - {\widetilde \mu}~{\widetilde P} =
 {{{\widetilde c}~{\widetilde P}^2 - 2~{\widetilde d}~{\widetilde P} + {\widetilde e}}\over{{\widetilde c}~{\widetilde e} - {\widetilde d}^2}}\label{g-pos}
\end{eqnarray}
where
\begin{eqnarray}
 &&{\widetilde c} \equiv c - a^T~Q^{-1}~a = \sum_{i,j\in J} C^{-1}_{ij}~\eta_i~\eta_j > 0\\
 &&{\widetilde d} \equiv d - b^T~Q^{-1}~a \label{dtilde} = \sum_{i,j\in J} C^{-1}_{ij}~\alpha_i~\eta_j\\
 &&{\widetilde e} \equiv e - b^T~Q^{-1}~b \label{etilde} = \sum_{i,j\in J} C^{-1}_{ij}~\alpha_i~\alpha_j > 0\\
 &&C^{-1}_{ij} = {1\over\xi_i^2}~\delta_{ij} - \sum_{A,B=1}^F {{\widetilde\Omega}_{iA}\over\xi_i^2}~{{\widetilde\Omega}_{iB}\over\xi_j^2}~Q^{-1}_{AB}
\end{eqnarray}
and $Q^{-1}_{AB}$ is the inverse of $Q_{AB}$.\footnote{\, Note that, because $C_{ij}$ has the factor model form, in this case the $N(J)\times N(J)$ matrix $D$ defined in (\ref{D}) {\em coincides} with the restriction of the inverse of the $N \times N$ matrix $C_{ij}$, $i,j\in 1,\dots,N$ to $i,j\in J$.} Note that if $\mu\geq 0$, then ${\widetilde\mu} <0$ assuming ${\widetilde P} = 0$.

{}From (\ref{mu.2}) for $\mu=0$ we have
\begin{eqnarray}
 &&{\widetilde P} = {\widetilde P}_* \equiv {{\widetilde e} \over {\widetilde d}} > 0\label{P.max.S}\\
 &&{\widetilde\mu} = -{1\over {\widetilde d}} < 0\label{tildemu.*}
\end{eqnarray}
We will now show that this corresponds to the weights ${\widetilde w}_i$ that maximize the Sharpe ratio: $S\rightarrow\mbox{max}$. (We will denote this maximal value of the Sharpe ratio via $S_{\rm{\scriptstyle{max}}}$.) These weights are given by
\begin{equation}\label{w.tilde}
 {\widetilde w}_i = \gamma\sum_{i=1}^N C^{-1}_{ij}~\alpha_j =
 {\gamma\over \xi^2_i}~\left(\alpha_i - \sum_{j = 1}^N {\alpha_j \over \xi^2_j}~\sum_{A,B = 1}^F {\widetilde\Omega}_{iA}~{\widetilde \Omega}_{jB}~Q^{-1}_{AB} \right)
\end{equation}
where $\gamma$ is fixed by (\ref{mod.w}). So we have (${\widetilde \eta}_i\equiv\mbox{sign}\left({\widetilde w}_i\right)$):
\begin{eqnarray}
 &&\left.{\widetilde P}~\right|_{S=S_{\rm{\scriptstyle{max}}}} = \sum_{i=1}^N \alpha_i~{\widetilde w}_i = \gamma~\left(e - \sum_{A,B=1}^F Q^{-1}_{AB}~b_A~b_B\right) = \gamma~{\widetilde e}\\
 &&1 = \sum_{i=1}^N {\widetilde \eta}_i~{\widetilde w}_i = \gamma\left(d - \sum_{A,B=1}^F Q^{-1}_{AB}~a_A~b_B\right) = \gamma~{\widetilde d}
\end{eqnarray}
and we have (\ref{P.max.S}).\footnote{\, The reason why $\mu=0$ corresponds to $S\rightarrow\mbox{max}$ can be understood by noting that for $\mu\neq 0$ the term proportional to $\mu$ in (\ref{g}) breaks the scaling property that $g \rightarrow \lambda^2 g$ under $w_i\rightarrow\lambda w_i$, ${\widetilde\mu}\rightarrow\lambda {\widetilde\mu}$ and ${\widetilde P}\rightarrow \lambda{\widetilde P}$.} Note that in this case generically $J^\prime$ is empty and ${\widetilde w}_i$ can only vanish ``accidentally" for those alphas such that the expression in the parenthesis in (\ref{w.tilde}) vanishes. In the following we will assume that such ``accidental" vanishings do not take place, {\em i.e.}, let us consider generic alpha configurations. Also, we will denote the values of ${\widetilde c}$, ${\widetilde d}$ and ${\widetilde e}$ at $S = S_{\rm{\scriptstyle{max}}}$ via ${\widetilde c}_* > 0$, ${\widetilde d}_* > 0$ and ${\widetilde e}_* > 0$. The second inequality follows from (\ref{tildemu.*}).

{}Next, note that we can focus on the values of ${\widetilde P}\geq {\widetilde P}_*$. Indeed, it makes no sense to consider ${\widetilde P}< {\widetilde P}_*$ as we have a solution with higher P\&L and higher Sharpe ratio, namely, the $S=S_{\rm{\scriptstyle{max}}}$ solution. Further, let us consider a solution with ${\widetilde P} = (1+\delta){\widetilde P}_*$, where $0 <\delta \ll 1$. Then $w_i = (1+\xi_i){\widetilde w}_i$, where $\left|\xi_i\right| \ll 1$ and we have taken into account that none of the ${\widetilde w}_i$ vanishes. We then have $\eta_i = {\widetilde\eta}_i$, and the values of ${\widetilde c}$, ${\widetilde d}$ and ${\widetilde e}$ for $w_i$ are the same as for ${\widetilde w}_i$. So we have
\begin{eqnarray}\label{mu.delta}
 &&\mu = {{\widetilde e}_*~\over {{\widetilde c}_*~{\widetilde e}_* - {\widetilde d}_*^2}}~\delta\\
 &&{\widetilde \mu} = -{{1+{\widetilde c}_*~\mu}\over{\widetilde d}_*}\\
 &&{\widetilde R}^2 = {{\widetilde e}_*\over{\widetilde d}_*^2}~\left(1 + {\widetilde c}_*~\mu~\delta\right)\label{vol}
\end{eqnarray}
Since for $\delta > 0$ the P\&L increases, the volatility cannot decrease as $\delta = 0$ corresponds to $S=S_{\rm{\scriptstyle{max}}}$. It then follows from (\ref{vol}) that $\mu > 0$ for $0<\delta \ll 1$ as we have ${\widetilde c}_* > 0$.

{}Note that $\mu = 0$ uniquely corresponds to the $S=S_{\rm{\scriptstyle{max}}}$ solution, so $\mu$ cannot become zero at any ${\widetilde P} > {\widetilde P}_*$. Therefore, as we smoothly increase ${\widetilde P}$, the only way $\mu$ can become negative is via a discontinuity in ${\widetilde c}$, ${\widetilde d}$ and ${\widetilde e}$, {\em i.e.}, for some ${\widetilde P} = {\widetilde P}_1 > {\widetilde P}_*$ we have $\mu = \mu_1 >0$ with one or more vanishing $w_i$, {\em i.e.}, $J^\prime \equiv J^\prime_1$ is not empty, while for ${\widetilde P} = (1 + \delta){\widetilde P}_1$, where $|\delta| \ll 1$, we have empty $J^\prime$ and $\mu = \mu_2(\delta) < 0$. This would imply that as $\delta\rightarrow 0$, we would have $w_i\rightarrow 0$ for $i\in J^\prime_1$ with non-vanishing $\mu = \mu_2(0) < 0$, which is impossible. (Note that ${\widetilde c}$, ${\widetilde d}$ and ${\widetilde e}$ are unchanged as $\delta\rightarrow 0$.) Also, $\mu > 0$ and $J^\prime$ is not empty as ${\widetilde P} \rightarrow \mbox{max}(\alpha_i)$. Finally, note that $\mu <0$ is compatible with ${\widetilde P} < {\widetilde P}_*$ as no $w_i$ have to vanish for $\mu$ to become 0 as ${\widetilde P}\rightarrow {\widetilde P}_*-$.

\subsection{$\mu < 0$ case}\label{sub5.4}

{}If $\mu < 0$, then we no longer have (\ref{eta1}) and (\ref{eta2}). Instead, from (\ref{globalmin}) it follows that $J^\prime$ is empty and $J=\{1,\dots,N\}$, {\em i.e.}, there are no vanishing $w_i$, and we have
\begin{eqnarray}
 &&\mu < 0\\
 &&w_i = {1\over\xi_i^2}~\left(\eta_i\left|\mu\right| + z_i\right),~~~i=1,\dots,N\\
 &&z_i \equiv - \left({\widetilde\mu}~\alpha_i + \sum_{A=1}^F {\widetilde \Omega}_{iA}~v_A\right),~~~i=1,\dots,N
\end{eqnarray}
It then follows that
\begin{eqnarray}
 &&\eta_i = \mbox{sign}\left(z_i\right),~~~i\in {\widetilde J}\\
 &&\forall i\in {\widetilde J}:~~~\left|z_i\right|\geq \left|\mu\right|
\end{eqnarray}
However, $\eta_i$ can be 1 or $-1$ for $i\not\in {\widetilde J}$, {\em i.e.}, it is undetermined.

{}This is because for $\mu < 0$ generically there are expected to be more than one local minima. This is problematic in two ways. First, it makes finding the global minimum more difficult due to the usual instabilities in any iterative procedure, especially when the number of local minima is large. Second, in the presence of a large number of local minima, the system is more prone to the out-of-sample instability as relatively small fluctuations in the off-diagonal elements of $C_{ij}$ can lead to an iterative algorithm jumping between local minima. However, as we discussed above, $\mu < 0$ luckily corresponds to the uninteresting range ${\widetilde P} < {\widetilde P}_*$.

\subsection{Conditions for Global Minimum}\label{sub5.5}

{}Above we gave the conditions for the global minimum:
\begin{eqnarray}\label{AJJ1}
 &&\sum_{i,j=1}^N C_{ij}~\left(w_i + \epsilon_i\right)~\left(w_j + \epsilon_j\right) \geq \sum_{i,j \in J} C_{ij}~w_i~w_j\\
 &&\sum_{i=1}^N \left|w_i + \epsilon_i\right| = 1\label{AJJ2}\\
 &&\sum_{i=1}^N \alpha_i~\left(w_i + \epsilon_i\right) = {\widetilde P}\label{AJJ3}
\end{eqnarray}
where $w_i$, $i\in J$ are determined using (\ref{J1}), while $w_i =0$, $i\in J^\prime$, and $\epsilon_i$ are arbitrary subject to (\ref{AJJ2}) and (\ref{AJJ3}). Above we discussed these conditions for arbitrary infinitesimal $\epsilon_i$, which give the conditions for a local minimum. Here we discuss the above conditions for non-infinitesimal $\epsilon_i$. We have
\begin{eqnarray}
 &&\sum_{i,j=1}^N C_{ij}~\epsilon_i~\epsilon_j + 2~\sum_{i,j \in J} C_{ij}~w_i~\epsilon_j + 2~\sum_{i\in J} \sum_{j\in J^\prime} C_{ij}~w_i~\epsilon_j \geq 0\\
 &&\sum_{i\in J} \left(\left|w_i + \epsilon_i\right| - \left|w_i\right|\right) + \sum_{i\in J^\prime} \left|\epsilon_i\right| = 0\label{AAJJ2}\\
 &&\sum_{i\in J} \alpha_i~\epsilon_i + \sum_{i\in J^\prime} \alpha_i~\epsilon_i = 0
\end{eqnarray}
which, taking into account (\ref{J1}) and (\ref{JJ4}), reduce to
\begin{equation}\label{AJJ4}
 \sum_{i,j=1}^N C_{ij}~\epsilon_i~\epsilon_j \geq 2~\mu~\left(\sum_{i\in J} \eta_i~\epsilon_i + \sum_{i\in J^\prime} \left|\epsilon_i\right|\right)
\end{equation}
Note that for infinitesimal $\epsilon_i$ the r.h.s. of (\ref{AJJ4}) vanishes due to (\ref{AAJJ2}), while the l.h.s. is positive due to positive-definiteness of $C_{ij}$. However, for non-infinitesimal $\epsilon_i$ we have nontrivial conditions.

{}Let us define $J_1\subset J$ and $J_2\subset J$ as follows:
\begin{eqnarray}
 &&\eta_i~(w_i +\epsilon_i) \geq 0,~~~i\in J_1\\
 &&\eta_i~(w_i +\epsilon_i) < 0,~~~i\in J_2
\end{eqnarray}
Then, from (\ref{AAJJ2}) we have
\begin{equation}
 \sum_{i\in J} \eta_i~\epsilon_i + \sum_{i\in J^\prime} \left|\epsilon_i\right| = -2~\sum_{i\in J_2} \left|w_i +\epsilon_i\right|
\end{equation}
and
\begin{equation}\label{globalmin1}
 \sum_{i,j=1}^N C_{ij}~\epsilon_i~\epsilon_j \geq -4~\mu~\sum_{i\in J_2} \left|w_i +\epsilon_i\right|
\end{equation}
This condition is always satisfied for $\mu \geq 0$, which implies that the solution to the local optimum conditions discussed above is automatically the global optimum. However, for $\mu < 0$ (\ref{globalmin1}) implies that a solution to the local optimum conditions is not guaranteed to be the global optimum. In fact, as we discuss in Subsection \ref{sub5.4}, for $\mu < 0$ we can have multiple local minima.

\subsection{General Case}

{}The methods employed in the context of the factor model can be used to tackle the case of a general covariance matrix with the understanding that, as we discuss below, convergence of an iterative procedure in the general case is tricky.

{}We want to solve (\ref{J1}) subject to (\ref{J2}) and (\ref{J3}) for a general positive-definite covariance matrix $C_{ij}$ (assuming $C_{ii} > 0$). From (\ref{J1}) we have
\begin{eqnarray}
 &&w_i = -{1\over C_{ii}}~\left(\mu~\eta_i + {\widetilde \mu}~\alpha_i + v_i\right),~~~i\in J\label{wi}\\
 &&v_i\equiv \sum_{j\in J(i)} C_{ij}~w_j,~~~i\in J\label{vi}\\
 &&J(i) \equiv J \setminus \{i\}
\end{eqnarray}
{\em i.e.}, $J(i)$ is defined as $J$ with the single-element subset $\{i\}$ subtracted, and $J$ (and $J^\prime$) are defined as above. Let
\begin{eqnarray}
 &&Q_{ij} \equiv {C_{ij} \over C_{jj}},~~~i,j \in J\\
 &&a_i \equiv \sum_{j \in J(i)} Q_{ij}~\eta_j = \sum_{j \in J} Q_{ij}~\eta_j - \eta_i,~~~i\in J\\
 &&b_i \equiv \sum_{j \in J(i)} Q_{ij}~\alpha_j = \sum_{j \in J} Q_{ij}~\alpha_j - \alpha_i,~~~i\in J\\
 &&{\widetilde a}_i \equiv {\eta_i \over C_{ii}},~~~i\in J\\
 &&{\widetilde b}_i \equiv {\alpha_i \over C_{ii}},~~~i\in J\\
 &&c\equiv \sum_{i\in J} {1\over C_{ii}}\\
 &&d\equiv \sum_{i\in J} {{\alpha_i~\eta_i}\over C_{ii}}\\
 &&e\equiv \sum_{i\in J} {{\alpha^2_i}\over C_{ii}}
\end{eqnarray}
Then, by plugging (\ref{wi}) into (\ref{vi}), (\ref{J2}) and (\ref{J3}), we obtain the following system of $N(J)+2$ equations for $N(J)+2$ unknowns $v_i$, $\mu$ and ${\widetilde \mu}$:\footnote{\, Recall the definition of $N(J)\equiv \left|J\right|$.}
\begin{eqnarray}
 &&\sum_{j\in J} Q_{ij}~v_j + a_i~\mu + b_i~{\widetilde \mu} = 0,~~~i\in J\\
 &&\sum_{i\in J} {\widetilde a}_i ~v_i + c~\mu + d~{\widetilde \mu} = 0\\
 &&\sum_{i\in J} {\widetilde b}_i ~v_i + d~\mu + e~{\widetilde \mu} = 0
\end{eqnarray}
Let ${\widetilde Q}_{ab}$, $a,b = 1,\dots,(N(J)+2)$ be the following $(N(J)+2) \times (N(J)+2)$ matrix:
\begin{eqnarray}
 &&{\widetilde Q}_{ij} \equiv Q_{ij},~~~i,j\in J\\
 &&{\widetilde Q}_{i,(N(J)+1)} \equiv a_i,~~~i\in J\\
 &&{\widetilde Q}_{(N(J)+1), i} \equiv {\widetilde a}_i,~~~i\in J\\
 &&{\widetilde Q}_{i,(N(J)+2)} \equiv b_i,~~~i\in J\\
 &&{\widetilde Q}_{(N(J)+2), i} \equiv {\widetilde b}_i,~~~i\in J\\
 &&{\widetilde Q}_{(N(J)+1),(N(J)+1)} \equiv c\\
 &&{\widetilde Q}_{(N(J)+1),(N(J)+2)} = {\widetilde Q}_{(N(J)+2),(N(J)+1)}\equiv d\\
 &&{\widetilde Q}_{(N(J)+2),(N(J)+2)} \equiv e
\end{eqnarray}
Note that $Q_{ab}$ is not symmetric. Also, let $x_a$ and $y_a$, $a=1,\dots,(N(J)+2)$ be the following $(N(J)+2)$-vectors:
\begin{eqnarray}
 &&x_i \equiv v_i,~~~i\in J\\
 &&x_{N(J)+1} \equiv \mu\\
 &&x_{N(J)+2} \equiv {\widetilde \mu}\\
 &&y_i = 0,~~~i\in J\\
 &&y_{N(J)+1} = -1\\
 &&y_{N(J)+2} = -{\widetilde P}
\end{eqnarray}
Then we have
\begin{equation}\label{NJ+2}
 x_a = \sum_{b=1}^{N(J)+2} {\widehat Q}_{ab}~y_b
\end{equation}
where ${\widehat Q}_{ab}$ is the matrix inverse\footnote{\, This inverse exists so long as $C_{ij}$ is invertible, which we assume to be the case.} to ${\widetilde Q}_{ab}$: ${\widehat Q} \equiv {\widetilde Q}^{-1}$.

{}Next, recalling that
$\eta_i~w_i > 0$, $i\in J$, using (\ref{wi}) and assuming $\mu \geq 0$ we have
\begin{eqnarray}
 &&\mu\geq 0\\
 &&\eta_i = -\mbox{sign}\left({\widetilde \mu}~\alpha_i + v_i\right),~~~i\in J\\
 &&\forall i\in J:~~~\left|{\widetilde \mu}~\alpha_i + v_i\right| > \mu\\
 &&\forall i\in J^\prime:~~~\left|{\widetilde \mu}~\alpha_i + v_i\right| \leq \mu
\end{eqnarray}
where the last condition follows from (\ref{globalmin}). The last three conditions determine $\eta_i$, $J$ and $J^\prime$ in terms of $v_i$, $\mu$ and ${\widetilde \mu}$, while (\ref{NJ+2}) solves $v_i$, $\mu$ and ${\widetilde \mu}$ given $\eta_i$, $J$ and $J^\prime$. So, one can attempt to solve the entire system iteratively as in the context of the factor model in Subsection \ref{sub5.2}. In this regard, the $N(J)$-vector $v_i$ here is analogous to the $F$-vector $v_A$ in Subsection \ref{sub5.2}. However, unless $N(J)\ll N$, the analogy with the factor model is only superficial and the convergence is not guaranteed.

\section{Linear Cost in Weight Optimization}\label{sec6}

{}Next, let us include linear cost in the weight optimization problem. Linear costs can be modeled by subtracting a linear penalty from the P\&L:
\begin{equation}
 P = I~\sum_{i=1}^N \alpha_i~w_i - L~D
\end{equation}
where $L$ includes all fixed trading costs (SEC fees, exchange fees, broker-dealer fees, {\em etc.}) and linear slippage.\footnote{Here for the sake of simplicity the linear slippage is assumed to be uniform across all alphas. This is not a critical assumption and can be relaxed, {\em e.g.}, by modifying the definition of $L_i$ below. In essence, this assumption is made to simplify the discussion of turnover reduction.} The linear cost assumes no impact, {\em i.e.}, trading does not affect the stock prices. Also, $D = I~T$ is the dollar amount traded, and $T$ is the turnover (so the turnover is defined as a percentage). Let $\tau_i$ be the turnovers corresponding to individual alphas $\alpha_i$. If we ignore turnover reduction resulting from combining alphas (or if the internal crossing is switched off), then
\begin{equation}
 T = \sum_{i=1}^N \tau_i~\left|w_i\right|
\end{equation}
However, with internal crossing turnover reduction can be substantial and needs to be taken into account. In \cite{SpMod} we proposed a model of turnover reduction, according to which when the number of alphas $N$ is large, the leading approximation (in the $1/N$ expansion) is given by
\begin{equation}\label{SM}
 T \approx  \rho_* \sum_{i=1}^N \tau_i~\left|w_i\right|
\end{equation}
where $0 < \rho_* \leq 1$ is the turnover reduction coefficient. Let us emphasize that this formula is expected to be a good approximation in the large $N$ limit (so long as the distribution of individual turnovers $\tau_i$ is not skewed) regardless of how $\rho_*$ is modeled. In \cite{SpMod} we also proposed a spectral model for estimating $\rho_*$ based on the correlation matrix $\Psi_{ij}$:
\begin{equation}
 \rho_* \approx {\psi^{(1)}\over{N\sqrt{N}}}~ \left|\sum_{i=1}^N {\widetilde V}^{(1)}_{i}\right|
\end{equation}
where $\psi^{(1)}$ is the largest eigenvalue of $\Psi_{ij}$ and ${\widetilde V}^{(1)}_{i}$ is the corresponding eigenvector normalized such that $\sum_{i=1}^N
\left({\widetilde V}^{(1)}_{i}\right)^2 = 1$.

{}We then have
\begin{equation}\label{L}
 P = I~\sum_{i = 1}^N \left(\alpha_i~w_i - L_i~\left|w_i\right|\right)
\end{equation}
where
\begin{equation}\label{L1}
 L_i \equiv L~\rho_*~\tau_i > 0
\end{equation}
We then follow the approach of Section \ref{sec5}, which is now modified as follows.

{}For the sake of brevity, we will not repeat the discussions of Section \ref{sec5} whose applicability is evident, and we will use the same notations as in Section \ref{sec5} and simply modify the corresponding definitions. Furthermore, we will assume the factor model of Subsection \ref{sub5.1}. We need to solve the following problem:
\begin{eqnarray}
 &&{P \over I} = \sum_{i=1}^N \left(\alpha_i~w_i - L_i~\left|w_i\right|\right) \equiv {\widetilde P} = \mbox{fixed}\\
 &&{R \over I} = \sqrt{\sum_{i,j=1}^N C_{ij}~w_i~w_j} \rightarrow \mbox{min}
\end{eqnarray}
subject to
\begin{equation}
 \sum_{i=1}^N |w_i| = 1
\end{equation}
This problem can be stated as follows:
\begin{eqnarray}
 &&g(w, \mu, {\widetilde \mu}) \equiv {1\over 2}~\sum_{i,j=1}^N C_{ij}~w_i~w_j +\nonumber \\
 &&~~~+ \mu\left(\sum_{i=1}^N \left|w_i\right| - 1\right) +
 {\widetilde \mu}\left(\sum_{i=1}^N \left(\alpha_i~w_i - L_i~\left|w_i\right|\right) - {\widetilde P}\right)
 \rightarrow \mbox{min}
\end{eqnarray}
We now have
\begin{eqnarray}\label{wv1}
 &&w_i = - {1\over \xi_i^2}~\left(\mu~\eta_i + {\widetilde \mu}~{\overline \alpha}_i + \sum_{A=1}^F {\widetilde \Omega}_{iA}~v_A\right),
 ~~~i\in J\\
 &&{\overline\alpha}_i \equiv \alpha_i - L_i~\eta_i,~~~i\in J
\end{eqnarray}
Recalling that we have $w_i~\eta_i > 0$, $i\in J$, and assuming that $\mu\geq {\widetilde \mu}~L_i$, $i\in J$ we get
\begin{eqnarray}\label{mu.cond.L.2}
 &&\mu \geq {\widetilde \mu}~L_i,~~~i = 1,\dots, N\\
 &&\eta_i = -\mbox{sign}\left({\widetilde\mu}~\alpha_i + \sum_{A=1}^F {\widetilde \Omega}_{iA}~v_A\right),~~~i\in J\label{eta1L}\\
 &&\forall i\in J:~~~\left|{\widetilde\mu}~\alpha_i + \sum_{A=1}^F {\widetilde \Omega}_{iA}~v_A\right| > \mu - {\widetilde\mu}~L_i\label{eta2L}\\
 &&\forall i\in J^\prime:~~~\left|{\widetilde\mu}~\alpha_i + \sum_{A=1}^F {\widetilde \Omega}_{iA}~v_A\right| \leq \mu - {\widetilde\mu}~L_i\label{eta3L}
\end{eqnarray}
Note that $\mu\geq {\widetilde \mu}~L_i$, $i\in J^\prime$ follows from (\ref{eta3L}).\footnote{\, The global minimum conditions in the presence of linear costs are discussed in Subsection \ref{sub6.1}, where we also derive the above conditions for a local minimum.}

{}The rest of the discussion is identical to that in Subsection \ref{sub5.2} with $\alpha_i$ replaced by ${\overline \alpha_i}$ in all the remaining definitions. We still have (\ref{mu.2}), (\ref{mutilde.2}) and (\ref{g-pos}), so for $\mu\geq 0$ we still have ${\widetilde\mu} < 0$, which implies that the condition (\ref{mu.cond.L.2}) is automatically satisfied for $\mu \geq 0$. {\em I.e.}, the condition for the existence of the unique minimum with the linear costs ($\mu\geq {\widetilde \mu}~L_i$) is not as strong as in the absence of costs ($\mu \geq 0$). Furthermore, $\mu = 0$ still corresponds to maximizing the Sharpe ratio $S\rightarrow \mbox{max}$ with linear costs \cite{AlphaOpt}. Let the corresponding ${\widetilde P}\equiv {\widetilde P}_*$. Then we can restrict to ${\widetilde P}\geq {\widetilde P}_*$, for which we have $\mu\geq 0$, which means that the results of Subsection \ref{sub5.2} apply.

{}Here the following remark is in order. Because the alphas $\alpha_i$, $i\in J^\prime$ are no longer traded, we can drop such alphas, if any, recompute $\rho_*$ in (\ref{L1}) using the corresponding correlation matrix $\Psi^\prime_{ij} \equiv \Psi_{ij}\left.\right|_{i,j\in J}$, recompute $w_i$ using such $\rho_*$ and repeat this procedure until the subset $J$ based on which $\rho_*$ is computed is the same as the subset for which $w_i\not=0$, where $w_i$ are computed based on such $\rho_*$.\footnote{\, When $N$ is large, this procedure is stable and convergent as $\rho_*$ does not change much with $N$ (see \cite{SpMod}).}

\subsection{Global Minimum Conditions with Linear Costs}\label{sub6.1}

{}The discussion in Subsection \ref{sub5.5} is modified as follows in the presence of linear costs. We have
\begin{eqnarray}\label{BJJ1}
 &&\sum_{i,j=1}^N C_{ij}~\left(w_i + \epsilon_i\right)~\left(w_j + \epsilon_j\right) \geq \sum_{i,j \in J} C_{ij}~w_i~w_j\\
 &&\sum_{i=1}^N \left|w_i + \epsilon_i\right| = 1\label{BJJ2}\\
 &&\sum_{i=1}^N \left(\alpha_i~\left(w_i + \epsilon_i\right) - L_i~ \left|w_i + \epsilon_i\right|\right) = {\widetilde P}\label{BJJ3}
\end{eqnarray}
where $w_i$, $i\in J$ are determined via
\begin{equation}\label{BB}
 \sum_{j\in J} C_{ij}~w_j = -\mu~\eta_i - {\widetilde \mu}~\left(\alpha_i - L_i~\eta_i\right),~~~i \in J
\end{equation}
while $w_i =0$, $i\in J^\prime$. We have
\begin{eqnarray}
 &&\sum_{i,j=1}^N C_{ij}~\epsilon_i~\epsilon_j + 2~\sum_{i,j \in J} C_{ij}~w_i~\epsilon_j + 2~\sum_{i\in J} \sum_{j\in J^\prime} C_{ij}~w_i~\epsilon_j \geq 0\\
 &&\sum_{i\in J} \left(\left|w_i + \epsilon_i\right| - \left|w_i\right|\right) + \sum_{i\in J^\prime} \left|\epsilon_i\right| = 0\label{BBJJ2}\\
 &&\sum_{i\in J} \left(\alpha_i~\epsilon_i - L_i~\left(\left|w_i + \epsilon_i\right| - \left|w_i\right|\right)\right) + \sum_{i\in J^\prime} \left(\alpha_i~\epsilon_i - L_i~\left|\epsilon_i\right|\right) = 0
\end{eqnarray}
which, taking into account (\ref{BB}), reduce to
\begin{eqnarray}
 &&\sum_{i,j=1}^N C_{ij}~\epsilon_i~\epsilon_j + 2~\sum_{j\in J^\prime} \left(\sum_{i \in J} C_{ij}~w_i~\epsilon_j + \left(\mu - {\widetilde \mu}~L_j\right)~\left|\epsilon_j\right| + {\widetilde \mu}~\alpha_j~\epsilon_j\right) \geq \nonumber\\
 &&~~~\geq -4~\sum_{i\in J_2} \left(\mu - {\widetilde \mu}~L_i\right)~\left|w_i + \epsilon_i\right| \label{BJJ4}
\end{eqnarray}
where, as above, $J_2$ is defined as follows:
\begin{equation}
 \eta_i~(w_i +\epsilon_i) < 0,~~~i\in J_2 \subset J
\end{equation}
For infinitesimal $\epsilon_i$ we have empty $J_2$ and (\ref{BJJ4}) reduces to
\begin{equation}
 \sum_{j\in J^\prime} \left(\sum_{i \in J} C_{ij}~w_i~\epsilon_j + \left(\mu - {\widetilde \mu}~L_j\right)~\left|\epsilon_j\right| + {\widetilde \mu}~\alpha_j~\epsilon_j\right)
 \geq 0
\end{equation}
which gives
\begin{equation}
 \forall j\in J^\prime:~~~\left|\sum_{i \in J} C_{ij}~w_i + {\widetilde \mu}~\alpha_j\right| \leq \mu - {\widetilde \mu}~L_j
\end{equation}
which in turn gives (\ref{eta3L}) in the factor model context.

{}For non-infinitesimal $\epsilon_i$ we have the following condition:
\begin{equation}
 \sum_{i,j=1}^N C_{ij}~\epsilon_i~\epsilon_j \geq -4~\sum_{i\in J_2} \left(\mu - {\widetilde \mu}~L_i\right)~\left|w_i + \epsilon_i\right|
\end{equation}
which is always satisfied if
\begin{equation}\label{mumutildeL}
 \mu \geq {\widetilde \mu}~L_i,~~~i\in J
\end{equation}
In this case the solution to the local optimum conditions discussed above is automatically the global optimum. However, if (\ref{mumutildeL}) is not satisfied, then a solution to the local optimum conditions is not guaranteed to be the global optimum. In fact, in this case we can have multiple local minima.

\section{Impact in Weight Optimization}\label{sec7}

{}Finally, let us discuss the effect of impact, {\em i.e.}, nonlinear costs, on weight optimization. Generally, introducing nonlinear impact makes the weight optimization problem computationally more challenging and requires introduction of approximation methods.

{}One way of modeling trading costs is to introduce linear and nonlinear terms:
\begin{equation}
 P = I~\sum_{i=1}^N \alpha_i~w_i - L~D - {1\over n}~Q~D^n
\end{equation}
where $D = I~T$ is the dollar amount traded, and $T$ is the turnover. If we model turnover using (\ref{SM}), then we have
\begin{equation}\label{impact}
 P = I~\sum_{i=1}^N \left(\alpha_i~w_i - L_i~\left|w_i\right|\right) - {1\over n}~{\widetilde Q}~\left(\sum_{i=1}^N \tau_i~\left|w_i\right|\right)^n
\end{equation}
where the modulus accounts for the possibility of some $w_i$ being negative, and ${\widetilde Q}$ is defined as follows
\begin{equation}
 {\widetilde Q}\equiv Q~(I~\rho_*)^n
\end{equation}
For general fractional $n$, which would have to be measured empirically, the weight optimization problem would have to be solved numerically.

{}First, note that if individual turnovers $\tau_i\equiv\tau$ are identical, then the nonlinear cost contribution into $P$ is independent of $w_i$ as we have (\ref{mod.w}). In this case, it simply shifts ${\widetilde P}$ by a constant and the problem can be solved exactly as in the previous section.\footnote{\, In fact, in this case the contribution of the linear cost also shifts ${\widetilde P}$ by a constant.} If $\tau_i$ are not all identical, then we need to solve the following problem:
\begin{eqnarray}
 &&g(w, \mu, {\widetilde \mu}) \equiv {1\over 2}~\sum_{i,j=1}^N C_{ij}~w_i~w_j + \mu~\left(\sum_{i=1}^N \left|w_i\right| - 1\right) + \nonumber \\
 &&~~~+
 {\widetilde \mu}~\left[\sum_{i=1}^N \left(\alpha_i~w_i - L_i~\left|w_i\right|\right) -{1\over n}~{\widetilde Q}^\prime~\left(\sum_{i=1}^N \tau_i~\left|w_i\right|\right)^n - {\widetilde P}\right]\\
 &&g(w, \mu, {\widetilde \mu}) \rightarrow \mbox{min}
\end{eqnarray}
where
\begin{equation}
 {\widetilde Q}^\prime \equiv {{\widetilde Q}\over I}
\end{equation}
Here one can use successive iterations to deal with the nonlinear term and various stability issues associated with convergence must be addressed. A simpler approach is, following \cite{AlphaOpt}, to note that the key role of the nonlinear term is to model portfolio capacity\footnote{\label{capacity}\, By this we mean the value of the investment level $I = I_*$ for which the P\&L $P_{\rm{\scriptstyle{opt}}}(I)$ is maximized, where for any given $I$ P\&L $P_{\rm{\scriptstyle{opt}}}(I)$ is computed for the optimized weights $w_i$. When only linear cost is present, capacity is unbounded. When nonlinear cost is included, $I_*$ is finite.} via its dependence on $I$, not its detailed structure in terms of individual alphas. In this regard, the following approximation is a reasonable way of simplifying the problem. Let
\begin{eqnarray}
 &&{\overline \tau}\equiv {1\over N}\sum_{i=1} \tau_i\\
 &&{\widetilde \tau}_i\equiv \tau_i - {\overline \tau}
\end{eqnarray}
If the distribution of ${\widetilde \tau}_i$ has a small standard deviation, then we can use the following approximation:
\begin{eqnarray}
 &&g(w, \mu, {\widetilde \mu}) \approx {1\over 2}~\sum_{i,j=1}^N C_{ij}~w_i~w_j + \mu~\left(\sum_{i=1}^N \left|w_i\right| - 1\right) + \nonumber \\
 &&+
 {\widetilde \mu}~\left[\sum_{i=1}^N \left(\alpha_i~w_i - L_i~\left|w_i\right|\right) - {1\over n}~{\widetilde Q}^\prime~\left({\overline \tau}^n + n~{\overline \tau}^{n-1}~\sum_{i=1}^N {\widetilde \tau}_i~\left|w_i\right|\right) - {\widetilde P}\right]
\end{eqnarray}
The objective function can be rewritten as
\begin{eqnarray}
 &&g(w, \mu, {\widetilde \mu}) \approx {1\over 2}~\sum_{i,j=1}^N C_{ij}~w_i~w_j + \mu~\left(\sum_{i=1}^N \left|w_i\right| - 1\right) + \nonumber \\
 &&~~~+
 {\widetilde \mu}~\left(\sum_{i=1}^N \left(\alpha_i~w_i - {\widetilde L}_i~\left|w_i\right|\right) - {\widetilde P}^\prime\right)
\end{eqnarray}
where
\begin{eqnarray}
 &&{\widetilde L}_i \equiv L_i + {\widetilde Q}^\prime~{\overline \tau}^{n-1}~\tau_i = L_i + Q~\rho_*^n~I^{n-1}~{\overline \tau}^{n-1}~\tau_i \label{Leff}\\
 &&{\widetilde P}^\prime \equiv {\widetilde P} + {1\over n}~{\widetilde Q}^\prime~{\overline \tau}^n
\end{eqnarray}
{\em I.e.}, in this approximation the effect of the nonlinear term reduces to increasing the linear slippage and shifting ${\widetilde P}$, and this problem we can solve as in the previous section. Note, however, that the ``effective" linear cost ${\widetilde L}_i$ now depends on the investment level $I$ via (\ref{Leff}), which now controls capacity. Thus, for $I$ such that
\begin{equation}
 {\widetilde L}_i \geq \left|\alpha_i\right|
\end{equation}
the P\&L cannot be positive.\footnote{\, The restriction on $I$ is even more severe, because this condition tells us that ${\widetilde P}^\prime$ cannot be positive, and for ${\widetilde P}$ to be positive, ${\widetilde L}_i$ would have to be even lower. At the end one is interested in determining capacity $I_*$, for which the optimized P\&L is maximized.}

\section{Comments}\label{sec8}

{}Above, among other things, we discussed optimization where one maximizes P\&L subject to a lower bound on the Sharpe ratio.
Theoretically this is a perfectly sound optimization problem. However, in practice it needs to be amended for the following reason.
Realized return and expected return generally are vastly different. Because of this, the
realized Sharpe ratio and the expected Sharpe ratio generally are also vastly different. For this
reason, the condition $S\geq S_{\rm{\scriptstyle{min}}}$ is impractical in terms of the realized Sharpe ratio as there
is no natural way of setting realized $S_{\rm{\scriptstyle{min}}}$, {\em i.e.}, the lower bound on the realized Sharpe ratio.
However, unlike the Sharpe ratio, portfolio risk $R$ is more stable and it makes sense
to replace the condition $S\geq S_{\rm{\scriptstyle{min}}}$ by $R\leq R_{\rm{\scriptstyle{max}}}$. {\em I.e.}, one now has the following
optimization problem: $P \rightarrow \mbox{max}$, $R\leq R_{\rm{\scriptstyle{max}}}$, where the condition on $R$ can now
be treated as a condition on expected risk. Happily, the actual solution to the
latter optimization problem is the same as that discussed in Sections \ref{sec5}, \ref{sec6} and \ref{sec7},
where now one specifies an upper bound on risk as opposed to
a lower bound on the Sharpe ratio. The same applies to our discussion in Subsection \ref{sub3.3}.

{}One of our main points above is that using factor models for alpha streams substantially reduces the number of iterations in the optimization problem and renders it practically tractable. Using a constructed covariance matrix as opposed to the one computed based on the alpha time series also has the advantage that -- assuming the factor covariance matrix is not computed using the very same time series -- the factor model covariance matrix is expected to be more stable out-of-sample. Also, the number of nonzero eigenvalues in the computed covariance matrix is limited to $M\ll N$ (where $M+1$ is the number of observations -- see Section \ref{sec.2}), {\em i.e.}, the number of risk factors (which are essentially based on principal components of the computed covariance matrix) is limited to $M$, whereas the constructed factor model covariance matrix can have many more risk factors as long as the factor covariance matrix $\Phi_{AB}$ is constructed based on stable data. Factor models for alpha streams is discussed in more detail in \cite{AlphaFM}.


\begin{thebibliography}{99}

\bibitem{HF1} T. Schneeweis, R. Spurgin, and D. McCarthy,
``Survivor Bias in Commodity Trading Advisor Performance",
J. Futures Markets, 1996, 16(7), 757-772.

\bibitem{HF2} C. Ackerman, R. McEnally and D. Revenscraft,
``The Performance of Hedge Funds: Risk, Return and Incentives",
Journal of Finance, 1999, 54(3), 833-874.

\bibitem{HF3} S.J. Brown, W. Goetzmann and R.G. Ibbotson,
``Offshore Hedge Funds: Survival and Performance, 1989-1995",
Journal of Business, 1999, 72(1), 91-117.

\bibitem{HF4} F.R. Edwards and J. Liew,
``Managed Commodity Funds",
Journal of Futures Markets, 1999, 19(4), 377-411.

\bibitem{HF5} F.R. Edwards and J. Liew,
``Hedge Funds versus Managed Futures as Asset Classes",
Journal of Derivatives, 1999, 6(4), 45-64.

\bibitem{HF6} W. Fung and D. Hsieh,
``A Primer on Hedge Funds",
Journal of Empirical Finance, 1999, 6(3), 309-331.

\bibitem{HF7} B. Liang,
``On the Performance of Hedge Funds",
Financial Analysts Journal, 1999, 55(4), 72-85.

\bibitem{HF8} V. Agarwal and N.Y. Naik,
``On Taking the ``Alternative" Route: The Risks, Rewards, and Performance Persistence of Hedge Funds",
Journal of Alternative Investments, 2000, 2(4), 6-23.

\bibitem{HF9} V. Agarwal and N.Y. Naik,
``Multi-Period Performance Persistence Analysis of Hedge Funds Source",
Journal of Financial and Quantitative Analysis, 2000, 35(3), 327-342.

\bibitem{HF10} W. Fung and D. Hsieh,
``Performance Characteristics of Hedge Funds and Commodity Funds: Natural vs. Spurious Biases",
Journal of Financial and Quantitative Analysis, 2000, 35(3), 291-307.

\bibitem{HF11} B. Liang,
``Hedge Funds: The Living and the Dead",
Journal of Financial and Quantitative Analysis, 2000, 35(3), 309-326.

\bibitem{HF12} C.S. Asness, R.J. Krail, and J.M. Liew,
``Do Hedge Funds Hedge?",
Journal of Portfolio Management, 2001, 28(1), 6-19.

\bibitem{HF13} F.R. Edwards and M.O. Caglayan,
``Hedge Fund and Commodity Fund Investments in Bull and Bear Markets",
Journal of Portfolio Management, 2001, 27(4), 97-108.

\bibitem{HF14} W. Fung and D. Hsieh,
``The Risk in Hedge Fund Strategies: Theory and Evidence from Trend Followers",
Review of Financial Studies, 2001, 14(2), 313-341.

\bibitem{HF15} B. Liang,
``Hedge Fund Performance: 1990-1999",
Financial Analysts Journal, 2001, 57(1), 11-18.

\bibitem{HF16} A.W. Lo,
``Risk Management For Hedge Funds: Introduction and Overview",
Financial Analysis Journal, 2001, 57(6), 16-33.

\bibitem{HF17} C. Brooks and H.M. Kat,
``The Statistical Properties of Hedge Fund Index Returns and Their Implications for Investors",
Journal of Alternative Investments, 2002, 5(2), 26-44.

\bibitem{HF18} D.-L. Kao,
``Battle for Alphas: Hedge Funds versus Long-Only Portfolios",
Financial Analysts Journal, 2002, 58(2), 16-36.

\bibitem{HF19} G. Amin and H. Kat,
``Stocks, Bonds and Hedge Funds: Not a Free Lunch!",
Journal of Portfolio Management, 2003, 29(4), 113-120.

\bibitem{HF20} N. Chan, M. Getmansky, S.M. Haas and A.W. Lo,
``Systemic Risk and Hedge Funds",
published in: Carey, M. and Stulz, R.M., eds.,
``The Risks of Financial Institutions" (University of Chicago Press, 2006), Chapter 6, 235-338.

\bibitem{OD} Z. Kakushadze and J.K.-S. Liew,
``Is It Possible to OD on Alpha?", The Journal of Alternative Investments (forthcoming); http://ssrn.com/abstract=2419415.

\bibitem{PO1} H. Markowitz,
``Portfolio selection",
Journal of Finance, 1952, 7(1), 77-91.

\bibitem{PO2} A. Charnes and W.W. Cooper,
``Programming with linear fractional functionals",
Naval Research Logistics Quarterly, 1962, 9(3-4), 181-186.

\bibitem{PO3} W.F. Sharpe,
``Mutual fund performance",
Journal of Business, 1966, 39(1), 119-138.

\bibitem{PO4} R.C. Merton,
``Lifetime portfolio selection under uncertainty: the continuous time case",
The Review of Economics and Statistics, 1969, 51(3), 247-257.

\bibitem{PO5} S. Schaible,
``Parameter-free convex equivalent and dual programs of fractional programming problems",
Zeitschrift f\"ur Operations Research, 1974, 18(5), 187-196.

\bibitem{PO6} M. Magill and G. Constantinides,
``Portfolio selection with transactions costs",
J. Econom. Theory, 1976, 13(2), 245-263.

\bibitem{PO7} A.F. Perold,
``Large-scale portfolio optimization",
Management Science, 1984, 30(10), 1143-1160.

\bibitem{PO8} M. Davis and A. Norman,
``Portfolio selection with transaction costs",
Math. Oper. Res., 1990, 15(4), 676-713.

\bibitem{PO9} B. Dumas and E. Luciano,
``An exact solution to a dynamic portfolio choice problem under transaction costs",
The Journal of Finance, 1991, 46(2), 577-595.

\bibitem{PO10} C. J. Adcock and N. Meade
``A simple algorithm to incorporate transactions costs in quadratic optimization",
European Journal of Operational Research, 1994, 79(1), 85-94.

\bibitem{PO11} S. Shreve and H.M. Soner,
``Optimal investment and consumption with transaction costs",
Ann. Appl. Probab., 1994, 4(3), 609-692.

\bibitem{PO12} D. Bienstock,
``Computational study of a family of mixed-integer quadratic programming problems",
Mathematical Programming, 1996, 74(2), 121-140.

\bibitem{PO13} J. Cvitani\'{c} and I. Karatzas,
``Hedging and portfolio optimization under transaction costs: a martingale approach",
Math. Finance, 1996, 6(2), 133-165.

\bibitem{PO14} A. Yoshimoto,
``The mean-variance approach to portfolio optimization subject to transaction costs",
J. Operations Research Soc. of Japan, 1996, 39(1), 99-117.

\bibitem{PO15} C. Atkinson, S.R. Pliska and P. Wilmott,
``Portfolio management with transaction costs",
Proc. Roy. Soc. London Ser. A, 1997, 453(1958), 551-562.

\bibitem{PO16} D. Bertsimas, C. Darnell and R. Soucy,
``Portfolio construction through mixed-integer programming at Grantham, Mayo, Van Otterloo and Company",
Interfaces, 1999, 29(1), 49-66.

\bibitem{PO17} A. Cadenillas and S. R. Pliska,
``Optimal trading of a security when there are taxes and transaction costs",
Finance and Stochastics, 1999, 3(2), 137-165.

\bibitem{PO18} T.-J. Chang, N. Meade, J.E. Beasley and Y.M. Sharaiha,
``Heuristics for cardinality constrained portfolio optimisation",
Computers and Operations Research, 2000, 27(13), 1271-1302.

\bibitem{PO19} H. Kellerer, R. Mansini and M.G. Speranza,
``Selecting portfolios with fixed costs and minimum transaction lots",
Annals of Operations Research, 2000, 99(1-4), 287-304.

\bibitem{PO20} R.T. Rockafellar and S. Uryasev,
``Optimization of conditional value-at-risk",
Journal of Risk, 2000, 2(3), 21-41.

\bibitem{PO21} J. Gondzio and R. Kouwenberg,
``High-performance computing for asset-liability management",
Operations Research, 2001, 49(6), 879-891.

\bibitem{PO22} H. Konno and A. Wijayanayake,
``Portfolio optimization problem under concave transaction costs and minimal transaction unit constraints",
Mathematical Programming, 2001, 89(2), 233-250.

\bibitem{PO23} S. Mokkhavesa and C. Atkinson,
``Perturbation solution of optimal portfolio theory with transaction costs for any utility function",
IMA J. Manag. Math., 2002, 13(2), 131-151.

\bibitem{PO24} O.L.V. Costa and A.C. Paiva,
``Robust portfolio selection using linear-matrix inequalities",
Journal of Economic Dynamics and Control, 2002, 26(6), 889-909.

\bibitem{PO25} F. Alizadeh and D. Goldfarb,
``Second-order cone programming",
Mathematical Programming, 2003, 95(1), 3-51.

\bibitem{PO26} M.J. Best and J. Hlouskova,
``Portfolio selection and transactions costs",
Computational Optimization and Applications, 2003, 24(1), 95-116.

\bibitem{PO27} K. Jane\v{c}ek and S. Shreve,
``Asymptotic analysis for optimal investment and consumption with transaction costs",
Finance Stoch., 2004, 8(2), 181-206.

\bibitem{PO28} M.S. Lobo, M. Fazel and S. Boyd,
``Portfolio optimization with linear and fixed transaction costs",
Annals of Operations Research, 2007, 152(1), 341-365.

\bibitem{PO29} R. Zagst and D. Kalin,
``Portfolio optimization under liquidity costs",
International Journal of Pure and Applied Mathematics, 2007, 39(2), 217-233.

\bibitem{PO30} M. Potaptchik, L. Tun\c{c}el and H. Wolkowicz,
``Large scale portfolio optimization with piecewise linear transaction costs",
Optimization Methods and Software, 2008, 23(6), 929-952.

\bibitem{PO31} E. Moro, J. Vicente, L.G. Moyano, A. Gerig, J.D. Farmer, G. Vaglica, F. Lillo and R.N. Mantegna,
``Market impact and trading profile of hidden orders in stock markets",
Physical Review E, 2009, 80, 066102.

\bibitem{PO32} J. Goodman and D.N. Ostrov,
``Balancing small transaction costs with loss of optimal allocation in dynamic stock trading strategies",
SIAM J. Appl. Math., 2010, 70(6), 1977-1998.

\bibitem{PO33} M. Bichuch,
``Asymptotic analysis for optimal investment in finite time with transaction costs",
SIAM J. Financial Math., 2012, 3(1), 433-458.

\bibitem{PO34} J.E. Mitchell and S. Braun,
``Rebalancing an investment portfolio in the presence of convex transaction costs, including market impact costs",
Optimization Methods and Software, 2013, 28(3), 523-542.

\bibitem{PO35} H. Soner and N. Touzi,
``Homogenization and asymptotics for small transaction costs",
SIAM Journal on Control and Optimization, 2013, 51(4), 2893-2921.

\bibitem{AlphaOpt} Z. Kakushadze, ``Combining Alpha Streams with Costs",
The Journal of Risk, 2015, 17(3), 57-78; http://ssrn.com/abstract=2438687; arXiv:1405.4716

\bibitem{SpMod} Z. Kakushadze, ``Spectral Model of Turnover Reduction",
SSRN Working Paper, http://ssrn.com/abstract=2427049; arXiv:1404.5050.

\bibitem{RJ} R. Rebonato and P. J\"ackel,
``The most general methodology to create a valid correlation matrix for risk management and option pricing purposes" (1999), http://ssrn.com/abstract=1969689 (December 7, 2011).

\bibitem{AlphaFM} Z. Kakushadze,
``Factor Models for Alpha Streams", The Journal of Investment Strategies, 2014, 4(1), 83-109;
http://ssrn.com/abstract=2449927; arXiv:1406.3396 [q-fin.PM].

\end{thebibliography}
\end{document}